\documentclass[a4paper,11pt]{article}

\usepackage{graphicx}
\usepackage{float}
\usepackage[T1]{fontenc}
\usepackage{epsfig}
\usepackage{color}
\usepackage{amsmath,jheppub,mathtools}
\usepackage{subfloat}
\usepackage{amsfonts}
\usepackage{braket}
\usepackage{cleveref}
\usepackage{epstopdf}
\usepackage{caption}
\usepackage{subcaption}
\usepackage{enumitem}
\usepackage[numbers]{natbib}
\usepackage[titletoc,toc,title]{appendix}
\usepackage{hyperref}
\hypersetup{
	colorlinks=true,
	linkcolor=blue,
	filecolor=red,      
	urlcolor=blue,
	citecolor=blue
} 

\title{\boldmath State Dependence of Krylov Complexity in $2d$ CFTs}

\author[a]{Arnab Kundu}
\author[a,b]{Vinay Malvimat}
\author[c]{Ritam Sinha}
\affiliation[a]{Theory Division, Saha Institute of Nuclear Physics,
	Homi Bhaba National Institute (HBNI),
	1/AF, Bidhannagar, Kolkata 700064, India.}
\affiliation[b]{Asia Pacific Center for Theoretical Physics, 77 Cheongam-ro, Nam-gu, Pohang-si, Gyeongsangbuk-do, 37673, Korea.}
\affiliation[c]{Department of Mathematics, King’s College London, Strand, London, WC2R 2LS, UK}

\emailAdd{arnab.kundu@saha.ac.in}
\emailAdd{vinay.malvimat@apctp.org}
\emailAdd{ritam.sinha@kcl.ac.uk}

\date{}

\abstract{We compute the Krylov Complexity of a light operator $\cO_L$ in an eigenstate of a $2d$ CFT at large central charge $c$. The eigenstate corresponds to a primary operator $\cO_H$ under the state-operator correspondence. We observe that the behaviour of K-complexity is different (either bounded or exponential) depending on whether the scaling dimension of $\cO_H$ is below or above the critical dimension $h_H=c/24$, marked by the $1st$ order Hawking-Page phase transition point in the dual $AdS_3$ geometry. Based on this feature, we hypothesize that the notions of operator growth and K-complexity for primary operators in $2d$ CFTs are closely related to the underlying entanglement structure of the state in which they are computed, thereby demonstrating explicitly their state-dependent nature. To provide further evidence for our hypothesis, we perform an analogous computation of K-complexity in a model of free massless scalar field theory in $2d$, and in the integrable $2d$ Ising CFT, where there is no such transition in the spectrum of states. }

\def\cO{\mathcal{O}}

\def\mF{\mathcal{F}}
\def\f{\frac}
\def\t{\tau}
\def\bz{\bar{z}}
\def\bw{\bar{w}}
\def\nn{\nonumber}
\def\a{\alpha}
\def\g{\gamma}
\def\s{\sigma}
\def\eps{\epsilon}
\newcommand\fnsep{\textsuperscript{,}}

\begin{document} 
	\maketitle
	
	\flushbottom

	\section{Introduction}

	The emergence of ergodicity and thermalization in many-body quantum systems is a challenging issue. While conceptual cornerstones {\it e.g.}~Eigenstate Thermalization Hypothesis (ETH) \cite{Srednicki:1994mfb} provide a crucial understanding for the emergence of these notions in closed isolated quantum systems, there are several alternative approaches and perspectives (see \cite{Gogolin:2015gts, DAlessio:2015qtq} for a review). Of these, the notion of operator growth in the operator Hilbert space of a quantum system with many degrees of freedom has recently been actively investigated (see {\it e.g.}~\cite{Parker:2018yvk,Avdoshkin:2022xuw, Camargo:2022rnt,Dymarsky:2021bjq, Dymarsky:2019elm, Avdoshkin:2019trj, Barbon:2019wsy, Khetrapal:2022dzy, Rabinovici:2021qqt, Rabinovici:2022beu} for a partial list of references). The idea of a simple and local operator growing into a complex non-local operator, under the Heisenberg time evolution, is believed to be responsible for ergodicity and thermalization in unitary quantum systems.

	This notion of growth is broadly related to the idea of complexity in quantum systems. Intuitively, complexity is a measure of the {\it difficulty}\footnote{More precisely, many such issues are connected to how hard it is to simulate the system in a computer. A fundamental notion of complexity is given in terms of the so-called Kolmogorov complexity of an object in the context of information theory.} in preparing a state in a quantum system, starting from a reference state. Currently, several inequivalent notions of complexity are being explored in the literature, which can be broadly classified into two qualitative kinds: One where the complexity of a state is probed
	%(compared to a given reference state)
	, and another in which the spread of a given operator is explored, as time evolves. Recently, several advances have been made in understanding better the notion of complexity, see {\it e.g.}~\cite{Balasubramanian:2022tpr, Caputa:2022eye, Bhattacharjee:2022lzy, Caputa:2022yju, Afrasiar:2022efk,Alishahiha:2022nhe} for quantum systems, \cite{Jefferson:2017sdb,Chapman:2018hou, Caceres:2019pgf, Chagnet:2021uvi} for quantum field theories, \cite{Stanford:2014jda, Susskind:2018pmk, Susskind:2018tei,Chattopadhyay:2023fob} for physics related to black holes, \cite{Chapman:2021jbh, Muck:2022xfc} for a recent review. At a conceptual level, these ideas are emerging as important ones across a very wide range of discrete and continuous systems, including {\it e.g.}~spin systems, quantum field theories and in quantum aspects of gravity. At the same time, continuum quantum field theories appear to pose certain challenges in attempts of directly implementing the ideas of discrete systems.

	On one hand, the idea of state complexity appears somewhat uneconomical, in that it requires several {\it ad hoc} parameters to be introduced for a working definition. Nonetheless, in recent literature related to quantum aspects of black holes, state complexity is widely believed to be important.\footnote{Note, however, that it is far from clear what precise geometric object captures the quantum state complexity of a strongly coupled large-$N$ gauge theory, see \cite{Belin:2021bga} for recent discussions on the non-uniqueness of choices in geometric observables.} On the other hand, the notion of operator complexity naively appears better suited for continuum quantum field theories as well as for quantum aspects of black holes, since it involves rather basic ingredients that seem readily generalizable from discrete to continuum descriptions. For example, given a {\it simple} and {\it local} operator $\cO$, its corresponding Heisenberg evolution, $e^{iHt} {\cO} e^{-iHt}$, spreads the operators through nested commutators $[H, [H, \ldots [H, \cO]]]$, where $H$ is the Hamiltonian of the system. To quantify the growth, one simply needs a definition of an inner product in the Hilbert space of operators and evaluate the spread of the operator wave functions in time.

	Based on these ideas, \cite{Parker:2018yvk} introduced the notion of Krylov complexity and subsequently conjectured a universal operator growth hypothesis for $1$d quantum systems. This proposal has been later explored further in various systems, including continuum quantum field theories in {\it e.g.}~\cite{Dymarsky:2021bjq, Dymarsky:2019elm, Khetrapal:2022dzy, Caputa:2021ori}. It has been noted in \cite{Dymarsky:2021bjq, Dymarsky:2019elm} that in the continuum limit, a strong kinematic, contact singularity in QFT yields a universal operator growth, despite the underlying dynamics of the QFT. While this could be indicative of a limitation of Krylov-complexity in QFT, it is worthwhile to revisit the issue.\footnote{Note that, already in \cite{Kar:2021nbm}, a proposal has been put forth regarding how one may be able to capture dynamical features in a QFT using K-complexity. However, this proposal is very different from ours, and in particular, does not make an explicit connection with a four-point function in CFT.} In this article, we propose that K-complexity indeed probes into the dynamics provided we opt for a general choice of operator norm. Earlier studies have relied on a thermal two-point function as an operator norm. Here we propose to generalize this to a correlation function in an excited state in the  spectrum of the theory.

	Our main framework is Conformal Field Theories (CFTs) in which the point above is crisply visible. For any CFT, in any dimension, the thermal two-point function of any two primary operators is fixed by conformal symmetry. Such a correlation function should not be expected to capture dynamical details of the corresponding CFT and, indeed, the conventional K-complexity always displays a kinematic universal behaviour of the so-called Lanczos coefficient. On the other hand, if we take the two-point correlation function of a given operator in a highly excited state (or a ``heavy'' eigenstate), it automatically becomes a four-point correlator which contains non-trivial dynamical information of OPE coefficient and fusion rules. We will show that this simple generalization indeed distinguishes between chaotic dynamics from integrable dynamics in a rather intriguing manner.

	The rest of the article is divided into the following sections. We begin with an introduction to the Krylov basis and Lanczos coefficients, thereafter briefly reviewing the existing results on K-complexity in CFTs, using the conventional norm.   We then introduce our proposal and demonstrate how dynamical features are encoded in the corresponding K-complexity, with several examples. We determine the K-complexity of a light operator in a large-c CFT and show that it exhibits a state dependent transition from an oscillating bounded growth to an exponential growth depending on the dimension of the heavy operator.  We have relegated several technical details to various appendices. 

	\section{A review of Krylov Complexity}
	In this section, we briefly review the Krylov algorithm to construct the Krylov basis in the operator Hilbert space of a quantum system \cite{VMR:2008vsg,Parker:2018yvk}. 
 We begin by considering a seed operator $\cO$ undergoing a Heisenberg time evolution,
\begin{equation}
	\mathcal{O}(t)=e^{i H t} \mathcal{O}(0) e^{-i H t}
\end{equation}
This time evolution can be re-expressed as an evolution of $\cO$ under a \emph{Liouvillian} super operator defined as,
\begin{align}
	\mathcal{O}(t)=e^{i H t} \mathcal{O}(0) e^{-i H t}\equiv e^{i \mathcal{L }t} \mathcal{O}(0)
\end{align}
where the Liouvillian operator is itself defined through the commutator with the Hamiltonian $[H,.]$, and its subsequent powers are defined by the following nested commutators,
\begin{align}
	\mathcal{H}_{\mathcal{O}}=\operatorname{span}\left\{\mathcal{L}^{n} \mathcal{O}\right\}_{n=0}^{+\infty}=\operatorname{span}\{\mathcal{O},[H, \mathcal{O}],[H,[H, \mathcal{O}]], \ldots\}
\end{align}
This linear span of operators forms an invariant subspace $\mathcal{H}_{\mathcal{O}}$. Thereby any operator within this subspace can be thought of as a vector in the linear vector space. Such a vector space, endowed  with a valid inner product, is called the \emph{Krylov subspace}. 

The vectors in this subspace are, however, not necessarily orthogonal. They can be made into an orthonormal basis using the Gram-Schmidt orthogonalisation procedure. The resulting set of operators form the \emph{Krylov basis}. Starting with the first basis element
\begin{align}
	\mid \mathcal{O}(t=0)):=\mid \mathcal{O}_{0}),
\end{align}
performing the Gram-Schmidt procedure leads to,\footnote{Usually, for the Wightman inner product, 
\begin{align}\label{2pt}
(	\mathcal{O}_1 \mid \mathcal{O}_2)=\operatorname{Tr}\left(\rho^{\frac{1}{2}} \mathcal{O}_1 ^{\dagger} \rho^{\frac{1}{2}} \mathcal{O}_2\right)
\end{align}
with $\rho=e^{-\beta H}$ being the thermal density matrix, with $\beta$ the inverse temperature and $H$ the system Hamiltonian. For this case, the coefficient $a_n=0$. However, we choose a non-standard inner product for evaluating correlation functions that makes $a_n\neq0$.},
\begin{align}
	\mid \tilde{\mathcal{O}}_{n})&=\mathcal{L} \mid \mathcal{O}_{n-1})-b_{n-1} \mid \mathcal{O}_{n-2})-a_{n-1}|{\cal O}_{n-1})\label{LactO}\\
	\mid \mathcal{O}_{n})&=b_{n}^{-1} \mid \tilde{\mathcal{O}}_{n}), \quad b_{n}=( \tilde{\mathcal{O}}_{n} \mid \tilde{\mathcal{O}}_{n})^{1 / 2}.\label{Basis2}
\end{align}
The coefficients $a_n,b_n$ appearing in the above expressions are known as the \emph{Lanczos coefficients} which are related to the matrix elements of the Liouvillian super-operator,
\begin{align}
	&( \mathcal{O}_{m}|\mathcal{L}|\mathcal{O}_m)=a_m\neq 0\\
	&( \mathcal{O}_{n-1}|\mathcal{L}|\mathcal{O}_{n}) =( \mathcal{O}_{n}|\mathcal{L}|\mathcal{O}_{n-1})=b_n
\end{align}	
The Liouvillian matrix in the Krylov basis has the following form,
\begin{align}
	L_{n m}:=(\mathcal{O}_{n}|\mathcal{L}| \mathcal{O}_{m})=\left(\begin{array}{ccccc}
		a_0 & b_{1} & 0 & 0 & \cdots \\
		b_{1} & a_1 & b_{2} & 0 & \cdots \\
		0 & b_{2} & a_2 & b_{3} & \cdots \\
		0 & 0 & b_{3} & a_3 & \ddots \\
		\vdots & \vdots & \vdots & \ddots & \ddots
	\end{array}\right)
\end{align}
For completeness, we outline the Toda chain technique for obtaining the Lanczos coefficients in Appendix \ref{Todasec}. Knowing these coefficients is important for computing the K-complexity. The detailed derivation of the technique can be found in \cite{Dymarsky:2019elm}.

The definition of a Krylov basis requires a valid inner product to be defined on the Krylov space, and the entire construction crucially depends on it.  For our purposes, we will define and use the following inner product,
\begin{align}\label{innerproductour}
	(\mathcal{O}_1 \mid \mathcal{O}_2)_\Psi=\f{\operatorname{Tr}\left(\rho_\Psi \, \mathcal{O}_1 ^{\dagger} \mathcal{O}_2\right)}{\operatorname{Tr}\rho_\Psi}
\end{align}
where $\rho_\Psi$ is some unnormalised density matrix corresponding to an excited state $|\Psi)$ in the theory, and which commutes with the Hamiltonian. This inner product satisfies all the norms for a valid inner product on the linear vector space of operators. Note that for this choice of the inner product, the $a_n$'s  do not vanish in general.
Any operator $\mathcal{O}(t)$ in the Hilbert space can now be expanded in terms of the Krylov basis elements $|\mathcal{O}_n)$,
\begin{align}\label{Ot}
	| \mathcal{O}(t))=\sum_{n} i^{n} \varphi_{n}(t) | \mathcal{O}_{n})
\end{align}
where $\varphi_{n}(t)$'s are complex coefficients. Solving the Heisenberg equation of motion $\partial_t\cO (t) = i[H,\cO(t)] $ produces a recursion relation for these coefficients,
\begin{align}\label{KWFEM}
	\partial_{t} \varphi_{n}(t)=b_{n} \varphi_{n-1}(t)-b_{n+1} \varphi_{n+1}(t)+i a_n\varphi_{n}(t)
\end{align}
with the boundary condition $\varphi_{-1}(t)=0$, along with $b_0=0$, and $\varphi_n(0)=\delta_{n,0}$. This equation is reminiscent of a particle hopping along a 1d spin chain.
The auto-correlation $C(t)$ is related to $\varphi_0(t)$ in the following way,
\begin{align}
	C_\Psi(t) \equiv(\mathcal{O}(0)|\mathcal{O}(t))_\Psi=\varphi_{0}(t)=\f{\operatorname{Tr}\left(\rho_\Psi \, \mathcal{O} ^{\dagger}(0) \mathcal{O}(t)\right)}{\operatorname{Tr}\rho_\Psi}
\end{align}
With all the definitions and set-up in place, we can now define the \emph{Krylov Complexity} (or \emph{K-complexity}) of an operator $\cO$ as its average size under time evolution,
\begin{align}
K_{\mathcal{O}}(t) =\left(\cO(t)|n|\cO(t)\right)=\sum_{n} n\left|\varphi_{n}(t)\right|^{2}
\end{align}
In what follows, we will compute the K-complexity of a primary operator in the excited state in several $2d$ CFTs.
% Thus, computing Krylov complexity involves determining the Lanczos coefficients, and $\varphi_{n}(t)$ through the recursion relation specified above.	
	\section{Krylov Complexity from four point function}
	
	\subsection{The setup}
	Our goal is to show the state dependence of Krylov complexity in $d\geq2$ QFTs. To achieve this, we will consider the simplest possible set-up of a $2d$ CFT. To compute the Krylov complexity for a primary operator $\cO$, we require its two-point function. We therefore begin with the computation of a 2-point function of a primary operator on a cylinder $S^1_L\times \mathbb{R}$, where $L$ is the compactification length of the spatial direction. We will quantize the theory on the $S^1_L$ with time flowing along the non-compact $\mathbb{R}$ direction. This allows us to insert state dependence into the computation by considering an excited state on the cylinder. Such a state is created by the insertion of a second conformal primary operator at the asymptotic infinities of the cylinder. Since the cylinder can be conformally mapped to the complex plane $\mathbb{C}$, the computation on the plane becomes that of a four-point function with two pairs of conformal primaries, out of which one pair from the asymptotic infinities gets mapped to respectively the origin and infinity of the planar geometry, under the conformal map. Since, our choice of quantisation on the cylinder maps to radial quantisation on the plane, we can treat the planar four-point function as a two-point function in an excited state.
As we will see, the Krylov complexity dependes on the state not just through the two-point function, but also through the inner product defined on the operator Hilbert space
%\footnote{ To compute the Krylov complexity in a pure excited state, we will need to introduce a new inner product.}. 
Let us begin by formally setting up the $2d$ CFT computation on the  $S^1_L\times \mathbb{R}$ geometry. We will denote the states at asymptotic infinity as $in$ and $out$ states on the cylinder in the following way,
\begin{equation}
 |\Psi_{in}\rangle = \lim_{T\rightarrow\infty}\cO_H(T)|0\rangle, ~~~\langle\Psi_{out}| = \lim_{T\rightarrow\infty}\langle 0|\cO_H(T)
\end{equation}
where $\cO_H$ is a primary operator with scaling dimension $\Delta_H=2h_H$. 
The auto-correlator of a different primary operator $\cO_L$ with scaling dimension $\Delta_L=2h_L$ is\footnote{The notation is chosen anticipating that $\cO_L$ and $\cO_H$ will denote light and heavy operators respectively in subsequent sections. However, for this discussion, $L$ and $H$ simply denote operators with different conformal dimensions.}.,
	\begin{align}
		C_\Psi(\t)=\langle{\cO_L(\t)} ~\cO_L(0)\rangle_{\Psi}=\frac{\langle\Psi_{out}|\cO_L(\t)\cO_L(0)|\Psi_{in}\rangle}{\langle\Psi_{out}|\Psi_{in}\rangle	}
	\end{align}
Note that we normalised the correlation function with the two-point function of the excited states since the pure density matrix $\rho_\Psi = |\Psi_{in}\rangle\langle\Psi_{out}|$ is un-normalised.
To compute $C_\Psi$, we conformally map the correlation functions on the cylinder to the plane via the conformal transformations, 
\begin{equation}
z=e^{\frac{2\pi}{L}w}, ~~\bar{z}=e^{\frac{2\pi}{L}\bar{w}}
\end{equation}
where $(z,\bar z)$ are the planar coordinates and $(w,\bar w)$ are the cylindrical coordinates. Moreover, we will assume that $(w,\bw)=(\t+i x,\t-ix)$, with $x\sim x+L$.
Under this map, the four-point function on the cylinder transforms as, 
\begin{align}
 &\frac{\langle\cO_H(w_1,\bw_1)\cO_L(w_2,\bw_2)\cO_L(w_3,\bw_3)\cO_H(w_4,\bw_4)\rangle}{\langle\cO_H (w_1,\bw_1)\cO_H(w_4,\bw_4)\rangle}=
\frac{\prod_{i=1}^4\bigg|\frac{dz_i}{dw_i}\bigg|^{h_i}\langle\cO_H(z_1)\cO_L(z_2)\cO_L(z_3)\cO_H(z_4)\rangle_{\mathbb{C}}\times ~ c.c.}{\prod_{i=1,4}\bigg|\frac{dz_i}{dw_i}\bigg|^{h_i}\langle\cO_H (z_1)\cO_H(z_4)\rangle_{\mathbb{C}}\times ~c.c.}\nn\\
&=\bigg(\frac{2\pi}{L}\bigg)^{2\Delta_L} |z_{14}|^{2\Delta_H}|z_2 ~z_3|^{\Delta_L}\langle{\cO_H(z_1)}\cO_L(z_2)\cO_L(z_3)\cO_H(z_4)\rangle_{\mathbb{C}}\times \langle{\cO_H(\bz_1)}\cO_L(\bz_2)\cO_L(\bz_3)\cO_H(\bz_4)\rangle_{\mathbb{C}}
\end{align}
where
% $(z,\b)$ are the planar coordinates, and $(w,\bw)=(\t+i x,\t-ix)$ are the coordinates on the cylinder, such that $x\sim x+L$, and 
$c.c.$ denotes the complex conjugate correlator. 
% Moreover, $\Delta_{L,H}=2h_{L,H}$ are the scaling dimensions of $\cO_{L,H}$
Conformal transformations on the plane reduce the four-point correlator to the function of a single cross-ratio (and its complex conjugate),
\begin{align}
 \zeta = \f{z_{12}z_{34}}{z_{13} z_{24}},~~~~ \bar{\zeta} = \f{\bz_{12} \bz_{34}}{\bz_{13}\bz_{24}}
\end{align}
To compute $C_\Psi$, we insert the operators at $w_1=T,w_2=\t,w_3=0,w_4=-T$ (and the same for $\bar{w}_i$) on the cylinder, while choosing all of them to lie on the same spatial slice $x=0$. On the plane, these correspond to,
\begin{align}
 z_1 =e^{\f{2\pi T}L},\hspace{.3cm} z_2 =e^{\f{2\pi \tau}L},\hspace{.3cm} z_3 =1,\hspace{.3cm} z_4 =e^{\f{-2\pi T}L}
 \label{insert}
\end{align}
In the $T\rightarrow\infty$ limit, the cross-ratios  become,
\begin{align}
 \zeta\sim e^{-\f{2\pi\t}L}, ~~~ \bar{\zeta}\sim e^{-\f{2\pi\t}L}
 \label{cross-largeT}
\end{align}
The auto-correlator $C_\Psi$ is therefore,
\begin{align}\label{S0tau}
C_\Psi(\t)&=\lim_{T\to\infty}\left(\frac{2\pi}{L}\right)^{2\Delta_L}(e^{\frac{2\pi}{L}T}-e^{-\frac{2\pi}{L}T})^{2\Delta_{H}}e^{\frac{2\pi\t}{L} \Delta_L}\langle{\cO_H(e^{\frac{2\pi}{L}T})}\cO_L(e^{\frac{2\pi}{L}\t})\cO_L(1)\cO_H(e^{-\frac{2\pi}{L}T})\rangle_{\mathbb{C}}\times ~c.c.\nn\\
&\approx \left(\frac{2\pi}{L}\right)^{2\Delta_L} |\zeta|^{-\Delta_L}\langle \cO_H(\infty)\cO_L(1/\zeta,1/\bar{\zeta})\cO_L(1)\cO_H(0)\rangle_{\mathbb{C}}
\end{align}
where we use the definition $\langle\cO(\infty)| = \lim_{\zeta_1,\bar{\zeta}_1\rightarrow\infty}|\zeta_1|^{2\Delta_H}\langle\cO(\zeta_1,\bar{\zeta}_1)|$. 
Using the conformal identity,
\begin{align}
 \langle \cO_H(\infty)\cO_L(1/\zeta,1/\bar{\zeta})\cO_L(1)\cO_H(0)\rangle_{\mathbb{C}} = |\zeta|^{2\Delta_L}\langle \cO_H(\infty)\cO_L(1)\cO_L(\zeta,\bar{\zeta})\cO_H(0)\rangle_{\mathbb{C}}
\end{align}
we can re-write the auto-correlation function in the standard notation,
\begin{align}
 C_\Psi(\t) \approx \left(\frac{2\pi}{L}\right)^{2\Delta_L} |\zeta|^{\Delta_L}\langle \cO_H(\infty)\cO_L(1)\cO_L(\zeta,\bar{\zeta})\cO_H(0)\rangle_{\mathbb{C}}
\end{align}
Such a four-point function on the plane can be computed exactly in free-theories, integrable models, as well as in certain limits of CFTs with large central charge $c$. For our purposes, we will first compute it in a large $c$ CFT, and thereafter compare and contrast our results with the integrable 2d CFTs such as the free scalar field theory and the Ising model. The objective will be to show the difference in behaviour of K-complexity as a function of the entanglement structure of the CFT eigenstates in these different theories.
\subsection{$2d$ CFT at large $c$}	
In this section we will compute the Lanczos growth and K-complexity of light operators in a large-$c$ CFT using their auto-correlator computed in eigenstates of varying entanglement structure, created through the insertion of light and heavy\footnote{Operators with scaling dimension $\Delta_L=2h_L$ such that $\f{h_L}c\sim\cO(1)$ in the $c\rightarrow\infty$ limit are called light, whereas the ones with scaling dimension $\Delta_H=2h_H$ such that $\f{h_H}c\gg 1$ in the $c\rightarrow\infty$ limit are called heavy.} primary states over the vacuum. 

\subsubsection*{The 4-point function at large $c$}
In a $2d$ CFT on $\mathbb{C}$ and with a large central charge, the leading large $c$ contribution to the four-point function of two light operators $O\equiv\mathcal{O}_L$, in a heavy state $\Psi\equiv\mathcal{O}_H$, follows just from the Virasoro vacuum block corresponding to the identity operator and its descendents \cite{Fitzpatrick:2014vua,Fitzpatrick:2015zha},
\begin{align}
\langle\mathcal{O}_H(\infty) \mathcal{O}_L(1)& \mathcal{O}_L(\zeta,\bar{\zeta})\mathcal{O}_H(0) \rangle=\mF (\zeta)\bar{\mF} (\bar{\zeta}) + \cO\left(\frac{1}{c}\right)
\end{align}
with the identity conformal block $\mF(\zeta)$ given by,
\begin{align}
\mF(\zeta)&=\gamma ^{2 h_L}\frac{ \zeta^{(\gamma -1) h_L}}{ \left(1-\zeta^{\gamma }\right)^{2 h_L}},\hspace{.3cm}\mbox{where}\hspace{.2cm}\gamma=\sqrt{1-24\frac{h_H}{c}}
\end{align}
and $\Delta_{L,H}=2h_{L,H}$ are the conformal dimensions of the light and heavy operators.
The auto-correlation function of $\cO_L$ on the cylinder $S^1_L\times \mathbb{R}$ is then computed using \eqref{cross-largeT} to be,
\begin{align}
C_\Psi(\t)=\frac{(\gamma\alpha)^{2\Delta_{L}}}{\sinh(\alpha\gamma \t)^{2\Delta_{L}}}
\label{light-corr}
\end{align}
where $\a=\frac{\pi}{L}$.

 To compute the K-complexity of $\cO_L$ starting from its auto-correlation function, we first need to determine the Lanczos coefficients $a_n$ and $b_n$. This is done using the method outlined in \cite{Dymarsky:2019elm}, where the problem of computing the Lanczos coefficients is mapped to an open Toda chain (see Appendix \ref{Todasec} for a review). The resulting differential equation, written in terms of Toda's tau function $\tau_n$, is solved subsequently using Hirota's bilinear technique. Below we will state the result for $\tau_n$ obtained by solving eq.\eqref{Toda}, and utilize the solution to compute the Lanczos coefficients $b_n$ and $a_n$ through  the relations given in eq.\eqref{bntoda} and eq.\eqref{antoda} .

Interestingly enough, $2d$ CFTs with a large $c$, defined at finite temperature and on a compact space, are holographically dual to a theory of Einstein gravity on $AdS_3$ spacetimes that have two separate phases in the bulk. These two phases, namely the \emph{thermal $AdS_3$} phase and the \emph{$AdS_3$ black hole} phase, are separated by a first-order \emph{Hawking Page phase transition}. Although our $CFT_2$ is defined at zero temperature, the insertion of a heavy operator in the vacuum puts the theory in an excited state, with a critical value for the scaling dimension $h_H = \f{c}{24}$, corresponding to the Hawking-Page phase transition in the dual gravity theory on $AdS_3$. For operators with scaling dimension below the critical value $h_H<\f{c}{24}$, the corresponding states are dual to an \emph{$AdS_3$ geometry with a conical deficit}, and therefore have an \emph{area-law entanglement} structure. This behaviour can be diagnosed in
the computation of the two-point function in a heavy eigenstate as outlined above. 

Let us get back to the computation of the Lanczos coefficients.
For the case of effectively \emph{light} external states with conformal dimension below the critical value, we use \eqref{light-corr} to get, 
 \begin{align}
	\tau_n(\t)	=\frac{G(n+2) G(n+2 \Delta_L +1)}{G(2 \Delta_L ) \Gamma (2 \Delta_L )^{n+1} }\frac{\left(\alpha\g\right)^{n(n+1)}}{\sinh \left(\alpha\g \t\right)^{(n+1) (n+2 \Delta_L )}}
\end{align}
where $G(m)$ denotes the Barnes gamma function.
The corresponding Lanczos coefficients are,
\begin{align}
	b_n^2=\frac{\left(\alpha\g\right)^2(n+1)(n+2\Delta_L)}{\sinh(\alpha\g\t)^2},\hspace{.3cm}a_n=-2\alpha\g(n+\Delta_L) \coth\left(\alpha\g \t\right)
	\label{lanc-light}
\end{align}

 On the other hand, for heavy operators with conformal dimension $h_H>\f{c}{24}$, i.e. above the critical value, the corresponding states are dual to \emph{black holes} in the Euclidean $AdS_3$ geometry. As a result, the states have have a \emph{volume-law entanglement} structure, as well as an emergent periodicity in the Euclidean time direction parameterized by $\beta_H = \f1{i \gamma} = \f1{\sqrt{\left(\f{24h_H}c-1\right)}}$. The auto-correlation function for these states becomes, 
\begin{align}\label{HHLLlargec}
C_\Psi(\t)=\frac{\left(\tilde{\alpha}\right)^{2\Delta_L}}{\sin\left(\tilde{\alpha} \t\right)^{2\Delta_L}}
\end{align}
where $\tilde{\alpha}=\alpha/\beta_H$.
One can compute Toda's tau function for this case in the exact same manner as above for the auto-correlator \eqref{HHLLlargec} to get,
\begin{align}
	 \tau_n(\t)	&=\frac{G(n+2) G(n+2 \Delta +1)}{G(2 \Delta ) \Gamma (2 \Delta )^{n+1} }\frac{(\tilde{\alpha}) ^{n(n+1)} }{\sin \left(\tilde{\alpha}  \t\right)^{(n+1) (n+2 \Delta )}}
\end{align}
The Lanczos coefficients corresponding to \eqref{HHLLlargec} are,
\begin{align}
b_n^2	&=\frac{\tilde{\alpha}^2(n+1)(n+2\Delta_L)}{\sin\left(\tilde{\alpha} \t\right)^2},\hspace{.3cm}a_n=-2\tilde{\alpha}(n+\Delta_L) \cot\left(\tilde{\alpha} \t\right)\label{lanc-heavy}
\end{align}
Under a simple analytic continuation of the time variable $\t$, the auto-correlation function in eq.\eqref{HHLLlargec} can be exactly identified with the thermal two-point function of a $2d$ CFT\footnote{For heavy enough states with $h_H>\f{c}{24}$, the four-point function in the $c\rightarrow\infty$ limit involving two heavy and two light operators behaves like a two-point function of light operators in the thermal state \cite{Fitzpatrick:2014vua,Fitzpatrick:2015zha}.}. This effectively proves that to leading order in large $c$, the heavy state indeed acts as a thermal state for probe light operators.

Despite the change in the entanglement structure of the states in which the auto-correlation function of $\cO_L$ is computed, the behaviour of the Lanczos coefficients remain linear in $n$ for asymptotically large $n$'s for $2d$ CFTs. This implies that, unlike in the $1d$ case, the linearity of the Lanczos coefficients in $n$ is not an adequate measure of chaotic behaviour in the system  \cite{Dymarsky:2021bjq}. As we will show below, the Krylov complexity of the CFT instead turns out to be more sensitive to this change in the entanglement structure of the state, and therefore proves to be a better diagnostic for quantum chaos in $2d$ CFTs .

\subsubsection*{K-Complexity for $2d$ CFTs at large $c$}

In this section, we will obtain the K-complexity of the primary operator $\cO_L$ for both the heavy and light external states. To do so, we will compute the amplitudes $\varphi_n$ from the differential equation \eqref{KWFEM}, using the Lanczos coefficients which we obtained previously. To solve \eqref{KWFEM}, it is imperative that we analytically continue to Lorentzian time $t$, while retaining a small ``cut-off" $\t_0$ in Euclidean space, such that $\t=\t_0+i \, t$. 
The auto-correlation function, following the analytic continuation becomes,
\begin{align}
C_\Psi ={\cal N} \varphi_0(t)&=\left(\frac{\alpha \gamma}{\sinh\left[\alpha \gamma\left(\t_0+i~ t\right)\right]}\right)^{2\Delta_L}
\label{Corr-large-c}
\end{align}
where ${\cal N}$ is a normalization constant.
This, along with the Lanczos coefficients \eqref{lanc-light}, are fed into the differential equation \eqref{KWFEM} to obtain the solution\footnote{Note that conventionally the Lanczos coefficients $b_n$s determined from the Toda chain method is  shifted  from $b_n\to b_{n-1}$ while solving for the amplitudes $\varphi_{n}$ using eq.\eqref{KWFEM} \cite{Parker:2018yvk,Avdoshkin:2022xuw,Dymarsky:2019elm,Dymarsky:2021bjq}.},
\begin{align}
\varphi_n(t)=\frac{1}{\text{csch}(\alpha \g \t_0)^{2\Delta_{L}}}\sqrt{\frac{\Gamma (n+2\Delta_L)}{\Gamma (2\Delta_{L}) \Gamma (n+1)} }\sin ^n(\alpha\g  t) \left(\sinh \left(\alpha\g  (\t_0 +i~t)\right)\right)^{-n-2\Delta_{L}}
\end{align}
Note that in the above expression we have fixed the normalization constant ${\cal N}=\text{csch}(\alpha \g \t_0)^{2\Delta_{L}}$ by demanding that the absolute square of the wave function sum to unity. We will follow the same procedure to fix the overall constant ${\cal N}$ in all the forthcoming cases.
It is easy to check that the above amplitude obeys the relation,
\begin{align}
\sum_{n=0}^{\infty}|\varphi_n(t)|^2=1
\end{align}
The K-complexity of $\cO_L$ in the light state is then,
\begin{align}
K_{L}(t)=\sum_{n=0}^{\infty}n~|\varphi_n(t)|^2=2\Delta_{L} \, \text{csch}^2(\alpha \g \t_0 ) \sin^2(\alpha \g t)
\end{align}
In the light state, therefore, the behaviour of K-complexity is \emph{bounded and oscillatory} despite the Lanczos coefficients being linear in $n$, with $n\rightarrow\infty$.

We will now do the exact same computation with the external states being heavy, i.e. for $h_H>\f{c}{24}$. Under the analytic continuation to real time, the correlation function of $\cO_L$ in the heavy state becomes,
\begin{align}\label{HHLLlargecLo}
C_\Psi(t)={\cal N}\varphi_0(t)=\frac{\left(\tilde{\alpha}\right)^{2\Delta_{L}}}{\sin\left(\tilde{\alpha} (\t_0+it)\right)^{2\Delta_{L}}}
\end{align}
Using the Lanczos coefficients \eqref{lanc-heavy} into \eqref{KWFEM}, we obtain the following solution for the amplitudes,
\begin{align}
\varphi_n=\frac{1}{(\text{csc}(\tilde{\alpha}  \t_0 ))^{2\Delta_{L}}}\sqrt{\frac{\Gamma (n+2\Delta_{L})}{\Gamma (2\Delta_{L}) \Gamma (n+1)} }\sinh ^n(\tilde{\alpha}  t) \sin\left(\tilde{\alpha}  (\t_0 +i t)\right) ^{-n-2\Delta_{L}}
\end{align}
Once again we have fixed the normalization constant ${\cal N}=\text{csc}(\tilde{\alpha}  \t_0 )^{2\Delta_{L}}$. Note that the above solution reduces to the family of  exact results obtained in \cite{Parker:2018yvk}  if we set the value of the cut-off to $\t_0=\frac{\pi \beta_H}{2\alpha}$,
\begin{align}
\varphi_n(t)=\sqrt{\frac{\Gamma (n+2\Delta_{L})}{\Gamma (2\Delta_{L})\Gamma\left(n+1\right) }}\frac{ \tanh ^n(\tilde{\alpha}  t)}{(\cosh (\tilde{\alpha} t))^{2\Delta_{L}}}
\end{align}
The corresponding K-complexity of $\cO_L$ in the heavy state is thus given by,
\begin{align}
K_H(t)=\sum_{n=0}^{\infty}n|\varphi_n(t)|^2=2\Delta_{L} \, \text{csc}^2(\tilde{\alpha}  \t_0 ) \sinh ^2(\tilde{\alpha}  t)
\end{align}
At large times, therefore, the K-complexity grows \emph{exponentially} fast,
\begin{align}
K_H(t)&\sim 2\Delta_{L} \,  \text{csc}^2(\tilde{\alpha}  \t_0) e^{2\tilde{\alpha} t}
\end{align}
This behaviour is in complete agreement with the behaviour observed for any other CFT in $d\geq2$ in the thermal state\footnote{Similar results for K-complexity have been obtained in \cite{Avdoshkin:2022xuw} in the black hole and thermal $AdS$ backgrounds in the holographic context. In fact, if we assume that the heavy eigenstates obey an ETH-like statement \cite{Lashkari:2016vgj}, then our results can also be viewed as being the dual computation on the $CFT_2$ side of \cite{Avdoshkin:2022xuw} in the $AdS_3$ bulk. We thank Anatoly Dymarsky for very helpful communication on this issue.}\fnsep\footnote{ For earlier work on similar results from Holography, see {\it e.g.}~\cite{Susskind:2019ddc}. We thank Julian Sonner for bringing this to our attention.}. However, it is in stark contrast with the behaviour observed above for light states. Our conclusion, therefore, is that \emph{K-complexity is a state dependent entity in higher dimensional CFTs, where it specifically depends on the entanglement structure of the state in which it is computed}.

\begin{figure}[H]
	\centering
	\begin{subfigure}{0.45\textwidth}
		\includegraphics[scale=0.4]{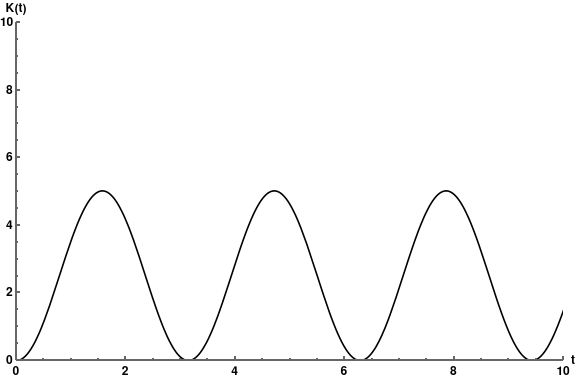}
		\caption{K-Compexity $K_L(t)$ vs t for LLLL .}
	\end{subfigure}
	\hfill
	\begin{subfigure}{0.45\textwidth}
		\centering
		\includegraphics[scale=0.4]{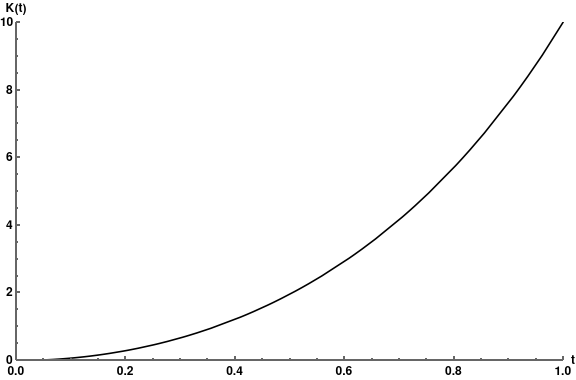}
		\caption{K-Compexity $K_H(t)$ vs t for HLLH.}
		\label{}
	\end{subfigure}
	\caption{ K-Complexity below (left) and above (right) the threshold  dimension of the heavy operator $h_H=c/24$ for the values of parameters $\alpha =1$, $\tau_0=0.98$, $\Delta_{L}=5$.}
	\label{Large-cHHLLAmpl2}
\end{figure}

% First, if we assume that an ETH-like statement\footnote{We thank Anatoly Dymarsky for suggesting this possibility. Note that, for us, ETH is an assumption. Also note that we are completely oblivious to the off-diagonal terms and their behaviour in this assumption, which are sometimes thought to contain more non-trivial physics related to ETH, compared to the diagonal answer.} holds for a class of CFT operators and CFT states, then $\langle h | {\cal O}(t) {\cal O}(0)| h\rangle$, where $h$ corresponds to the dimension of the operator that creates the state, is expected to behave as a thermal correlator with an effective temperature. Depending on the state, {\it i.e.}~$h$, the corresponding $2$-point function obeys a non-thermal or a thermal-like behaviour. While the former should correspond to a thermal AdS-background, the latter is expected to correspond to a BTZ-like geometry. This picture is also consistent with the behaviour obtained in \cite{Avdoshkin:2022xuw}. 

We leave this section with the following remark.
The {\it transition} from a \emph{bounded} to an \emph{exponential} growth of K-complexity is reminiscent of a similar transition in the OTOCs observed in \cite{Anous:2019yku}\footnote{We thank Julian Sonner for pointing this out to us.} where it was computed respectively in heavy and light eigenstates (based on a six-point function computation in \cite{Banerjee:2016qca}). Perhaps this alludes to a connection between scrambling of operators with that of entropy production in chaotic quantum systems. Nonetheless, K-complexity appears to be more economical in that it involves a four-point function, as opposed to a six-point function for OTOCs in \cite{Anous:2019yku}.

\subsection{Integrable $2d$ CFTs}
The large $c$ CFT in $2d$ is unique in that the nature of entanglement of a state corresponding to a primary operator $\cO_H$, transitions from an area law to a volume law structure exactly at the critical point $h_H=\f{c}{24}$. It thus makes for the perfect candidate to test our hypothesis regarding the state dependent nature of K-complexity. In this section, we will provide further evidence for our hypothesis by computing K-complexity in certain integrable $2d$ CFTs which does not have such a transition at any point along their spectrum. Therefore, if the hypothesis is correct, the nature of K-complexity in such theories will remain unchanged no matter how heavy the external state becomes, in comparison with the light operator in question. 

\subsubsection{The free massless scalar field theory in $2d$}
The simplest example of an integrable CFT is a free massless scalar field theory in $2d$. We can consider two classes of primary operators in this theory: 1. $\cO_L\equiv\partial \phi$ with scaling dimension $\Delta_L=1$, and 2. the vertex operator $\Psi\equiv e^{i k\phi}$ with scaling dimension $\Delta_\Psi = \f{k^2}{4\pi}$\footnote{This scaling dimension follows from the following convention for the action,
\begin{align}
 S = \int\,d^2x~\f12\left(\partial_\mu\phi\partial^\mu\phi\right)
\end{align}
}. The auto-correlation function of $\cO_L$ in the state corresponding to $\Psi$ is,
\begin{align}
 C_\Psi(\t)=\langle{\cO_L(\t)} ~\cO_L(0)\rangle_{\Psi}=\frac{\langle\Psi_{out}|\cO_L(\t)\cO_L(0)|\Psi_{in}\rangle}{\langle\Psi_{out}|\Psi_{in}\rangle	}
\end{align}
Since the theory is free, all correlation functions can be computed using Wick contraction of the fundamental field $\phi$. 
The resulting correlation function is completely \emph{state independent} in the appropriate limits,
\begin{align}
 C_{\Psi}(\tau) &= \lim_{T\rightarrow\infty}\left(\f{2\pi}L\right)^{2}\left(z_2 ~z_3\right)\left(\f1{z_{12}z_{34}}+\f1{z_{13}z_{24}}-\f1{4\pi}\f1{z_{23}^2}\right)\nn\\
 &\approx -\f1{\pi}\left(\f{\a}{\sinh(\a~\tau)}\right)^2
\end{align}
On analytically continuing to Lorentzian time, the corresponding K-complexity is bounded and oscillating akin to $K_L(t)$ in the large $c$ CFT for $\cO_L$ in the light state. This is because  the above auto-correlation has the same form as that of a light state in the large-c CFT obtained in eq.\eqref{Corr-large-c}  upon choosing $\Delta_{L}=1$ and $\gamma=1$.

\subsubsection{The $2d$ Ising CFT}
The case of the free massless scalar field theory is somewhat trivial. We can test our hypothesis in a more non-trivial example of the $2d$ Ising CFT, which is an interacting integrable model. This theory has two-nontrivial primary operators, 1. the spin operator $\s$ with scaling dimension $\Delta_\s =2h_\sigma= 1/8$, and 2. the energy operator $\epsilon$ with scaling dimension $\Delta_\eps =2h_\eps= 1$.

\subsubsection*{The $\langle\s\s\s\s\rangle$ correlator}
We begin by considering the four-point correlation function involving the spin operator $\sigma$. In our notation : $\cO\equiv\Psi\equiv\sigma$. The four-point function of the $\sigma$ operators on the cylinder is,
\begin{align}
&C_\s\left(\tau\right) =\f{\langle{\s(w_1,\bw_1)}\s(w_2,\bw_2)\s(w_3,\bw_3)\s(w_4,\bw_4)\rangle_{S^1_L\times\mathbb{R}}}{\langle\s(w_1,\bw_1)\s(w_4,\bw_4)\rangle_{S^1_L\times\mathbb{R}}}\\
&=\lim_{T\rightarrow\infty}\left(\f{2\pi}L\right)^{2\Delta_\s}|z_{14}|^{2\Delta_\s}|z_2~z_3|^{\Delta_\s}\times\f12\bigg|\frac{z_{13} z_{24}}{z_{12} z_{41}z_{23}z_{34}}\bigg|^{2\Delta_\s}\left(|1+\sqrt{1-x}|+|1-\sqrt{1-x}|\right)\nn
\end{align}
Evaluating this expression with the insertions on the plane as specified in \eqref{insert}, we get
\begin{align}
C_\s(\tau)=\bigg(\frac{ \alpha e^{2\a\t}}{ \text{sinh}\left(\alpha \t\right)}\bigg)^{\frac{1}{4}}
\end{align}
This result matches with the light operator correlation in the light state with the identifications, $\Delta_L=1/8$ and $\g=1$ and an additional factor of $e^{\alpha\tau/2}$. The corresponding expression for $\tau_n$ is thus,
\begin{align}
\tau_n(\t)=\frac{G(n+\frac{5}{4}) G(n+2)}{G(\frac{1}{4})(\Gamma(\frac{1}{4}))^{n+1}}\frac{\alpha^{(n+1)(n+\frac{1}{4})}}{\left(\sinh\left(\alpha \t\right)\right)^{(n+1)(n+\frac{1}{4})}}\times e^{\f{(n+1)}2\tau}
\end{align}
The Lanczos coefficients that follow are,
\begin{align}
	b_n^2=\frac{\alpha^2(n+1)(n+\f14)}{\sinh^2(\alpha \t)},\hspace{.3cm}a_n=-2\alpha\left(n+\f18\right) \coth\left(\alpha \t\right)
	\label{lanc-lightIsing}
\end{align}
The $b_n(\t)$'s are again linear in $n$ in the asymptotic limit $n\rightarrow\infty$. However, the slope crucially depends on the distance between operator insertions on the cylinder in Euclidean time $\t$.
\subsubsection*{The $\langle\eps\s\s\eps\rangle$ correlator}
Consider now the mixed correlator involving the auto-correlation function of $\s$ in the state created by the energy operator $\epsilon$. Thus, we now have $\cO\equiv\sigma$ and $\Psi\equiv\epsilon$. The corresponding auto-correlation function on the cylinder geometry is \footnote{For the $\langle\eps\s\s\eps\rangle$ and  $\langle\eps\eps\eps\eps\rangle$ correlators on the complex plane, we follow the conventions in \cite{Mattis:1987mat} },
\begin{align}
&C_\eps\left(\tau\right) =\f{\langle{\eps(w_1,\bw_1)}\s(w_2,\bw_2)\s(w_3,\bw_3)\eps(w_4,\bw_4)\rangle_{S^1_L\times\mathbb{R}}}{\langle\eps(w_1,\bw_1)\eps(w_4,\bw_4)\rangle_{S^1_L\times\mathbb{R}}}\\
&=\lim_{T\rightarrow\infty}\left(\f{2\pi}L\right)^{2\Delta_\s}|z_{14}|^{2\Delta_\eps}|z_2~z_3|^{\Delta_\s}\times\bigg|\f1{z_{14}^{\Delta_\eps}z_{32}^{\Delta_\s}}\f{\left[\left(z_1 z_4 + z_2 z_3\right)-\f12\left(z_1 + z_4\right)\left(z_2 + z_3\right)\right]}{\sqrt{z_{12}z_{13}z_{24}z_{34}}}\bigg|^2\nn
\end{align}
Again, using the operator insertions at the points \eqref{insert}, the 
auto-correlation function is,
\begin{align}
C_\eps(\t)=\left(\f{\alpha}{\sinh\left(\alpha\tau\right)}\right)^{1/4}\left( \cosh(\alpha \t)\right)^2
\end{align}
The $\tau_n$'s  and hence the Lanczos coefficients for the above auto-correlations do not have an analytic solution in the closed form. Hence, we determine the $a_n$s and $b_n$s numerically. Both these coefficients start displaying a linear behaviour for relatively small values of $n$\footnote{This is also the case for free field theories in higher dimensions \cite{Dymarsky:2021bjq, Avdoshkin:2022xuw}. }, as can be seen in the plots below.

\begin{figure}[H]
	\centering
	\begin{subfigure}{0.45\textwidth}
		\centering
	\includegraphics[scale=0.35]{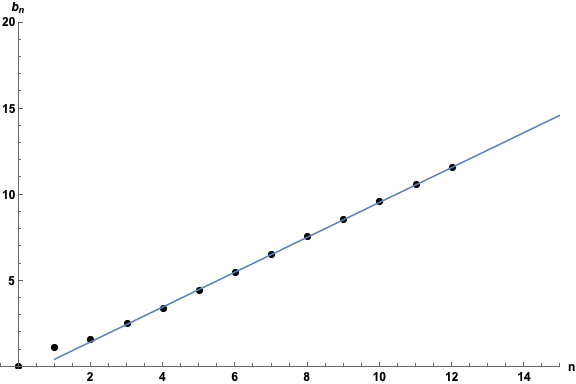}
	\caption{ $b_n$ vs n for the $ \sigma$ operator in the $\ket{\eps}$ state }
	\label{IsingHLLHbn}
		\label{}
	\end{subfigure}
	\hfill
\begin{subfigure}{0.45\textwidth}
	\centering
	\includegraphics[scale=0.35]{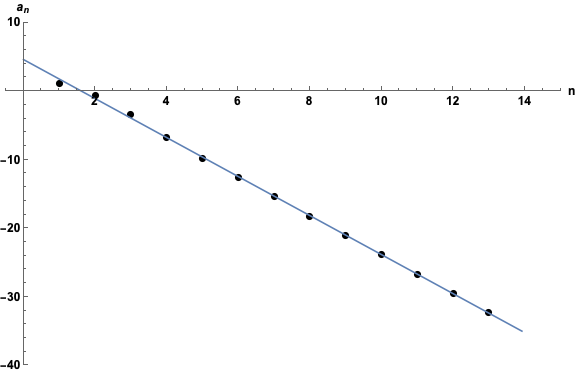}
	\caption{ $a_n$vs n  for  the $ \sigma$ operator in the $\ket{\eps}$ state }
	\label{IsingHLLHan}
	\label{}
\end{subfigure}
	\label{IsingHLLHLanczos}
	\caption{ Lanczos Coefficients $a_n, b_n$ for a light operator in heavy state in the Ising model example with  values of parameters $\alpha =1$, $\tau_0=0.88$, $\Delta_{H}=1$, $\Delta_{L}=1/8$.}
\end{figure}
\subsubsection*{The $\langle\eps\eps\eps\eps\rangle$ correlator}
We now consider the energy operator $\epsilon$ in a state created by its own insertion in the vacuum. In our notation this corresponds to $\cO\equiv\Psi\equiv\epsilon$. The required four point function in this case is given as follows \cite{Mattis:1987mat},
\begin{align}
	&C_\eps\left(\tau\right) =\f{\langle{\epsilon(w_1,\bw_1)}\epsilon(w_2,\bw_2)\epsilon(w_3,\bw_3)\epsilon(w_4,\bw_4)\rangle_{S^1_L\times\mathbb{R}}}{\langle\epsilon(w_1,\bw_1)\epsilon(w_4,\bw_4)\rangle_{S^1_L\times\mathbb{R}}}\nn\\
	&=\lim_{T\rightarrow\infty}\left(\f{2\pi}{L}\right)^{2\Delta_\eps}|z_{14}|^{2\Delta_\eps}|z_2 z_3|^{\Delta_\eps}\bigg|\frac{1}{z_{12}z_{34}}+\frac{1}{z_{14}z_{23}}-\frac{1}{z_{13}z_{24}}\bigg|^2
\end{align} 
Using the operator insertions at \eqref{insert}, we obtain the following expression for $C_\eps(\t)$,
\begin{align}
	C_\eps(\t)=\alpha^2\bigg[8\cosh\big(2\alpha\t\big)+\frac{1}{\sinh^2\big(\alpha \t\big)}\bigg]
\end{align}
Once again the auto-correlation is used to compute the sequence of Toda tau functions $\tau_n(\tau)$,
\begin{align}\label{tauLLLLIsing}
	\tau_n(\t)=&\frac{G(n+2) G(n+3)\alpha^{(n+1)(n+2)}}{\big(\sinh(\alpha \t)\big)^{(n+1)(n+2)}}F(n,\alpha \t)
\end{align}
where the function $F(n,\a\t)$ has an analytic form,
\begin{align}
	F(n,\alpha \t)=&g_0(n)+g_1(n)\cosh(2 \alpha \t)+g_2(n)\cosh(4 \alpha \t)\nn\\
	&+g_3(n)\cosh(2(n+1) \alpha \t)+g_4(n)\cosh(2(n+2) \alpha \t)
\end{align}
Here, $g_i(n)$'s are polynomials in $n$ of degree $\leq 4$. Their explicit forms are given by,
\begin{align}
	&g_0(n)=\frac{1}{2} \left(-3 n^4-18 n^3-35 n^2-24 n+6\right),\hspace{.3cm}g_1(n)=2 (n+1) (n+2) \left(n^2+3 n+1\right),\nonumber\\
	&g_2(n)=-\frac{1}{2} (n+1)^2 (n+2)^2,\hspace{.3cm}
	g_3(n)=-4 (n+2),\hspace{.3cm}
	g_4(n)=4 (n+1)
\end{align}
The corresponding Lanczos coefficients $b_n$ turn out to be,
\begin{align}
	b_n^2(\t)=\frac{(n+1)(n+2)\alpha^2}{\sinh(\alpha \t)^2}\frac{F(n+1,\alpha \t)~F(n-1,\alpha \t)}{F(n,\alpha \t)^2}
\end{align}
Similarly, the Lanczos coefficients $a_n$'s may be obtained by substituting $\tau_n(\tau)$ in eq.\eqref{tauLLLLIsing} into the relation \eqref{antoda},
\begin{align}
a_n(\t)	= -2\alpha \, (n+1) \coth (\alpha  \t)+\alpha \left(\frac{F'(n,\alpha  \t)}{F(n,\alpha \t)}-\frac{F'(n-1,\alpha  \t)}{F(n-1,\alpha  \t)}\right)
\end{align}	
where $F'$ denotes  the derivative of the function w.r.t the Euclidean time $\tau$.
\subsubsection*{K-complexity for the Ising CFT}
Having determined the Lanczos coefficients for the various correlators in the Ising model above, we will now use them to determine the behaviour of their corresponding K-complexity.

\subsubsection*{The $\langle\s\s\s\s\rangle$ correlator}
We first analytically continue the auto-correlation function $\tau_0(\tau)$ to complex time $\tau_0\rightarrow\tau_0 + it$.
\begin{equation}
	\varphi_0(t)=\frac{1}{{\cal N}}\bigg(\frac{ \alpha e^{2\alpha (\tau_0+i t)}}{ \text{sinh}\left(\alpha  (\tau_0+i t)\right)}\bigg)^{\frac{1}{4}}.
\end{equation}
We then substitute the above expression for $\varphi_{0}(t)$ and the Lanczos coefficients $a_n$ and $b_n$ obtained above with the cut-off set at $\tau=\tau_0$  in eq.\eqref{KWFEM} to solve for the amplitudes $\varphi_{n}$. 
%After appropriate normalization we obtained the following solution
\begin{align}
	\varphi_n(t)=\frac{1}{{\cal N}}\sqrt{\frac{\Gamma \left(n+\frac{1}{4}\right)}{\Gamma \left(\frac{5}{4}\right) \Gamma (n+1)}}\frac{2^n  e^{\alpha  (n+3) (\tau_0 +i t)} \sin ^n(\alpha  t)}{ \left(-1+e^{2 \alpha  (\tau_0 +i t)}\right)^{n+1} \left( e^{2 \alpha  (\tau_0 +i t)} \text{csch}[\alpha  (\tau_0 +i t)]\right)^{3/4}}.
\end{align}
where we have fixed the  being the cut-off dependent normalization constant ${\cal N}$ as earlier to be
\begin{align}
{\cal N}=	( e^{2 \alpha  \tau_0 } \text{csch}(\alpha  \tau_0))^{1/4}.
\end{align}
The corresponding K-complexity is then found to be,
\begin{align}
	K(t)=\frac{1}{4} \text{csch}^2(\alpha  \tau_0 ) \sin ^2(\alpha  t).
\end{align}
\begin{figure}[H]
	\centering
	\includegraphics[scale=0.46]{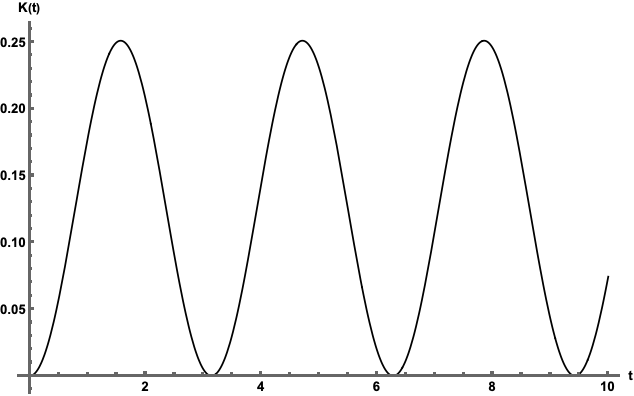}
	\caption{ K-complexity for a light operator ($\sigma$) in the light state $\ket{\sigma}$ in the Ising model example for the values of parameters $\alpha =1$, $\tau_0=0.88$, $\Delta_{\sigma}=1/8$.}
	\label{IsingLLLLAmpl2}
\end{figure}
%%%%%%%%%%%
\subsubsection*{The $\langle\eps\s\s\eps\rangle$ correlator}
The Lanczos coefficients and the amplitudes $\varphi_{n}(t)$ corresponding to the $\langle\eps\s\s\eps\rangle$ correlator are not analytically tractable. Therefore we resort to numerics and the plot for the time evolution of the K-complexity  is depicted in the figure below.
\begin{figure}[H]
		\centering
		\includegraphics[scale=0.46]{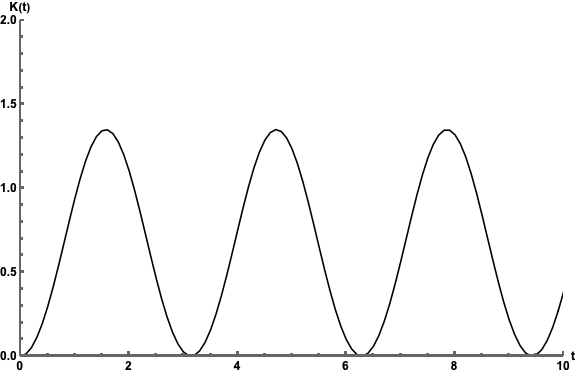}
	\caption{ K-complexity for a light operator ($\sigma$) in the heavy state $\ket{\eps}$ in the Ising model for the values of parameters $\alpha =1$, $\tau_0=0.88$, $\Delta_{\eps}=1/2$, $\Delta_{\sigma}=1/8$.}
	\label{IsingHLLHKTt}
\end{figure}

\subsubsection*{The $\langle\eps \eps \eps \eps\rangle$ correlator}
For the $\langle\eps \eps \eps \eps\rangle$ correlator on the other hand the Lanczos coefficients were analytically obtained previously. However, due to the complicated form of the Lanczos coefficients, the amplitudes and hence the K-Complexity are again not analytically solvable. Hence, we determine the K-complexity numerically as is displayed in the plot below.
\begin{figure}[H]
	\centering
	\includegraphics[scale=0.46]{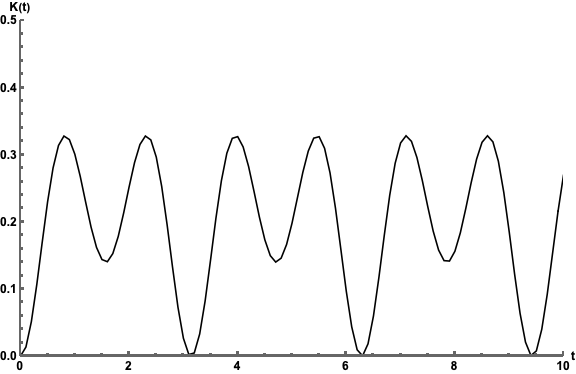}
	\caption{  K-complexity for the $\eps$ operator in $\ket{\eps}$  state  in the Ising model for the values of parameters $\alpha =1$, $\tau_0=0.88$, $\Delta_{\eps}=1/2$.}
	\label{IsingHHHHKt}
\end{figure}
To summarise, the K-complexity of a light operator is bounded and oscillatory, irrespective of whether the state in which it is computed is light or heavy, in both the 2d integrable CFTs involving the free massless scalar field theory and the Ising CFT. Thus, unlike in the large $c$ CFT example, for the integrable examples there is no transition of the K-complexity from a bounded and oscillatory behaviour in a light state, to an exponential behavior in a heavy state. It should be noted that all eigenstates in integrable quantum systems such as the ones discussed here, no matter how highly excited, exhibit an \emph{area law} entanglement structure. These results are then not only in accordance with our hypothesis, but provide further evidence in its favour.
\section{Discussions}
%%%%%%%%%%%

One of the primary focuses of this work has been to understand and make precise the role played by  four-point functions of two dimensional CFTs in the analysis of operator growth on the operator Hilbert space, using the tool of Krylov complexity. This seems intuitive since for a $CFT_2$ only four and higher point functions contain truly dynamical information about the system and therefore a distinction in operator growth between various quantum systems should be tied to such a dynamical correlation function, rather than a two-point function which is completely fixed kinematically\footnote{ Note that although we have restricted ourselves to 2d CFTs in the present article, the  Wightman inner product leading to an exponential growth of Krylov complexity (and linear growth of Lanczos coefficients ) has been demonstrated for various higher dimensional theories such as massless bosonic and fermionic free field theories \cite{Dymarsky:2019elm}. Hence, although there is no rigorous mathematical proof, the evidence from the above mentioned examples in \cite{Dymarsky:2019elm} suggest that such an universal exponential growth of K-Complexity exists as long as it is computed from the thermal two point function even in higher dimensional QFTs and as argued in \cite{Dymarsky:2021bjq, Dymarsky:2019elm, Avdoshkin:2022xuw}, this universality only senses the contact divergence of the correlator. In \cite{Avdoshkin:2022xuw}, the authors have described another intriguing feature of the Lanczos coefficient arising in higher dimensional free scalar field theory in the presence of mass, UV-cut off or compact support.}. In this article, we explicitly demonstrate the manner in which this intuition plays out, based on observations in various two-dimensional CFTs, where we carefully examine the structure of a four-point function in order to define an operator growth and a corresponding complexity.

One immediate conceptual consequence of our proposal is the {\it state-dependence} of K-complexity, which suggests that the corresponding dynamical feature is highly dependent on the region of the spectrum one is exploring. In this sense, the operator K-complexity mixes the notion of how {\it complex} the underlying dynamics is with how {\it complex} the state is. Intuitively, this is consistent with the recent studies on the detailed properties of the spectrum of chaotic quantum systems that includes states, such as the scar states (see \cite{Serbyn:2020wys, Moudgalya:2021xlu} for a review, {\it e.g.}~\cite{Banerjee:2020tgz, Biswas:2022env} for some recent developments, that do not thermalize. The latter fact makes it manifest that physical properties can crucially depend on the state as well. Furthermore, the analogue of K-Complexity in the state space  known as the  spread complexity was shown to exhibit revivals for such scar states in certain lattice models \cite{Bhattacharjee:2022qjw}. In our view, this makes K-complexity a more plausible candidate for capturing dynamical information, and one should not worry too much about the universal results that one obtains using a thermal two-point function. It further raises a possibility of connecting state-complexity with operator-complexity in some quantitative form \cite{Alishahiha:2022anw}. 

Another interesting observation we made from explicit computation is that the linearity of Lanczos coefficients does not necessarily translate into an exponential growth for the K-complexity. The standard folklore sometimes assumes otherwise. This is not a contradiction, since there is no theorem claiming the above statement, which we explicitly demonstrate in Appendix \ref{Hopsol}. In our analysis, the tunability of the slope of this linear behaviour played a crucial role. It will be very interesting to look for a potential upper bound on the slope and explore the corresponding implications, which we also hope to come back to in the recent future. 

There are several immediate applications and follow-ups of our work. For example for driven CFTs in two-dimensions (see {\it e.g}~\cite{Das:2021gts, Wen:2021mlv, Fan:2020orx, Wen:2020wee} for a partial list of recent activities along this direction), it is known that the four-point function (through the OTOC) indeed distinguishes the dynamics between {\it e.g.}~Ising CFT and a large-$c$ CFT, see {\it e.g.}~\cite{Das:2022jrr, Das:2022pez}. This is a framework particularly suited to further check our claims and establish whether the K-complexity is a good measure of thermal behaviour of the state and therefore a chaotic behaviour of the underlying dynamics. Moreover, it will be interesting to understand whether K-complexity can detect the first order phase transition from heating to non-heating phase, in this framework. More generally, it is also unclear to what extent complexity knows about phase transitions of a general order.

From a Holographic perspective, the key ingredients in computing the operator K-complexity are quite well-understood, and in principle a dual geometric description should exist (see \cite{Caputa:2022zsr, Rabinovici:2023yex} for recent works along this direction, and \cite{Susskind:2019ddc} for an earlier reference). We hope to visit this problem in the context of large c CFTs in the near future. Somewhat related to this are similar aspects in higher dimensional CFTs. For the latter, perhaps it is easier to find integrable examples, compared to non-trivial chaotic examples. Explorations along some of these directions are currently in order. 

Note that, four-point (as well as higher point) functions in CFTs make explicit and intriguing appearance in defining the so-called {\it early-time chaos} and provide corresponding bounds on the Lyapunov exponents \cite{Maldacena:2015waa}. While these bounds, especially their saturation, seem rooted in kinematic geometric factors \cite{Banerjee:2019vff, Banerjee:2018kwy, Banerjee:2018twd,Malvimat:2021itk,Malvimat:2022oue,Malvimat:2022fhd}, the physical picture nonetheless is deeply tied to gravitational wormholes and information transfer through them, see {\it e.g.} \cite{Gao:2016bin, Maldacena:2017axo, Kundu:2021nwp}. It will be very interesting to make some of these potential connections quantitative and precise.

\section*{Acknowledgements}
We thank Suman Das, Sabyasachi Maulik for initial collaboration on related topics; Debashis Banerjee, Diptarka Das, Shouvik Datta, Surbhi Khetrapal for various discussions and conversations related to this article; Pawel Caputa, Sumit Das, Anatoly Dymarsky, Julian Sonner for numerous comments and feedbacks on the manuscript. RS would like to thank SINP, Kolkata for hospitality during the initial stages of this work.
 AK is partially supported by CEFIPRA $6304-3$, DAE-BRNS 58/14/12/2021-BRNS and CRG/2021/004539 of Govt.~of India. RS is supported by the Royal Society-Newton International Fellowship, NIF/R1/221054-Royal Society. The work of VM was supported by the NRF grant funded by the Korea government (MSIT) (No. 2022R1A2C1003182) and by the Brain Pool program funded by the Ministry of Science and ICT through the National Research Foundation of Korea (RS-2023-00261799).

\appendix

\section{Lanczos coefficients through the Toda Heirarchy}\label{Todasec}
In this section, we will briefly describe the Toda hierarchy technique (see  \cite{Dymarsky:2019elm} for a detailed derivation) for obtaining the Lanczos coefficients $b_n$ and $a_n$. We can always associate a symmetric, positive definite matrix $G$ with the orthogonal inner product of the Krylov basis elements $|\cO_n)$. If the basis elements are made to dependent on (Euclidean) time, then so does $G\equiv G(\tau)$. The explicit time dependence of $G$ defines a geodesic flow in the space of symmetric positive-definite matrices, which in turn is captured by an open Toda chain described in terms of a Toda tau function $\tau_n(\tau)$. The open Toda chain equations can alternatively be written in the Hirota's bilinear form,
\begin{align}\label{Toda}
	\tau_n \ddot{\tau}_n-\dot{\tau}_n^2=\tau_{n+1} \tau_{n-1}, \quad \tau_{-1} \equiv 1
\end{align}
where the dots represent derivatives w.r.t. $\tau$. The auto-correlation function serves as a boundary condition for the Toda chain,
\begin{align}\label{t0ac}
\tau_{n=0}(\tau)\equiv C_\Psi(\tau)
\end{align}
along with $\tau_{-1}=1$. 
Together with these boundary conditions, the Toda chain \eqref{Toda} can be solved recursively, to obtain the sequence of $\tau_n$'s. Thereafter, the Lanczos coefficients $b_n$ and $a_n$ can be obtained from $\tau_n$ in the following way \cite{Dymarsky:2019elm},
\begin{align}
b_n&=\left.\frac{\tau_{n-1} \tau_{n+1}}{\tau_n^2}\right|_{\tau=\tau_0},\label{bntoda} \\
a_n&=\left.\frac{d}{d \tau}\log\bigg(\frac{\tau_n}{\tau_{n-1}}\bigg)\right|_{\tau=\tau_0} \label{antoda}.
\end{align}
where $\tau_0$ corresponds to the cut-off in the Euclidean time which we introduced to regulate the UV divergences.

\section{Solution of the hopping equation}\label{Hopsol}

Let us revisit one of the key equations that we have already described in the text (assume $a_n=0$),%
\begin{equation}
\partial_t \varphi_n = b_n \varphi_{n-1} - b_{n+1} \varphi_{n+1} \ . \label{hopeqn}
\end{equation}
We will now explore the broad feature of the amplitudes $\varphi_n$, given a linear spectrum of the Lanczos coefficients: $b_ n = \alpha n$, $\alpha \in {\mathbb R}$. Suppose we choose the ansatz that $\varphi_n(t) = e^{i \lambda t} q_n(t)$, such that $\varphi_n(0)=0$ except for $n=0$. This implies: $q_n(0)=0$, except $n=0$. Here $\{\lambda , q_n\} \in {\mathbb R}$. Substituting this ansatz in (\ref{hopeqn}) we obtain:
\begin{eqnarray}
i \lambda q_n(t) + \partial_t q_n(t) = \alpha n q_{n-1}(t) - \alpha(n+1) q_{n+1}(t) \ .
\end{eqnarray}
The above equation has no solution in our chosen range of various parameters. However, if we allow $\{q_n\}$ to be complex-valued, $q_n(t) = x_n(t) + i y_n(t)$, then the above equation splits into a coupled difference equation:
\begin{align}
 - \lambda y_n(t) &= \alpha n x_{n-1}(t) - \alpha(n+1) x_{n+1}(t) - \partial_t x_n(t) \ ,   \\
 \lambda x_n(t) &= \alpha n y_{n-1}(t) - \alpha(n+1) y_{n+1}(t) - \partial_t y_n(t) \ . 
\end{align}

This coupled system certainly has solutions. For example, taking $n \to \infty$ (such that $(n+1) \to n$), the difference equation can be transformed into a system of coupled differential equations which can be shown explicitly to have non-vanishing solutions. However, upon imposing the boundary condition $q_n(0)=0$ for $n\not =0$, the solution vanishes. So, the only non-vanishing component of the solution comes from $q_0$, {\it i.e.}~$\varphi_0$. Thus, with a linear growth of Lanczos coefficients, a bounded oscillatory behaviour of the corresponding K-complexity is certainly possible in this case.

This can be easily generalized for $a_n\not = 0$. In this case, the equation in (\ref{hopeqn}) has an additional $i a_n \varphi_n(t)$-term. Using the same ansatz as above, one obtains:
\begin{eqnarray}
\partial_t q_n(t) + i \left(\lambda - a_n \right) q_n (t) - b_n q_{n-1} (t) + b_{n+1} q_{n+1} (t) = 0 \ ,
\end{eqnarray}
which clearly admits non-trivial solutions when $\{q_n, a_n, b_n, \lambda\} \in {\mathbb R}$, by simply arranging $\lambda = a_n$. Once again, upon imposing the boundary condition $q_n(0)=0$ for $n\not =0$, leaves us with a non-vanishing $q_0$ which can finally yield an oscillatory K-complexity.  

These features explicitly demonstrate that even with a linear growth of Lanczos coefficients, it is possible to obtain a bounded operator amplitude. While this does not guarantee a bounded K-complexity, it is certainly allowed. Note that in our examples above, this qualitatively means that the operator becomes highly localized at large times, since only $q_0$ contribute. This is reminiscent of an integrable behaviour.


\begin{thebibliography}{22}
%\cite{Srednicki:1994mfb}
\bibitem{Srednicki:1994mfb}
M.~Srednicki,
``Chaos and Quantum Thermalization,''
doi:10.1103/PhysRevE.50.888
[\href{https://arxiv.org/abs/cond-mat/9403051}{{\tt arXiv:cond-mat/9403051 [cond-mat]}}].
%716 citations counted in INSPIRE as of 12 Dec 2022

%\cite{Gogolin:2015gts}
\bibitem{Gogolin:2015gts}
C.~Gogolin and J.~Eisert,
``Equilibration, thermalisation, and the emergence of statistical mechanics in closed quantum systems,''
Rept. Prog. Phys. \textbf{79}, no.5, 056001 (2016)
doi:10.1088/0034-4885/79/5/056001
[\href{https://arxiv.org/abs/1503.07538}{{\tt arXiv:1503.07538 [quant-ph]}}].
%399 citations counted in INSPIRE as of 27 Feb 2023

%\cite{DAlessio:2015qtq}
\bibitem{DAlessio:2015qtq}
L.~D'Alessio, Y.~Kafri, A.~Polkovnikov and M.~Rigol,
``From quantum chaos and eigenstate thermalization to statistical mechanics and thermodynamics,''
Adv. Phys. \textbf{65}, no.3, 239-362 (2016)
doi:10.1080/00018732.2016.1198134
[\href{https://arxiv.org/abs/1509.06411}{{\tt arXiv:1509.06411 [cond-mat.stat-mech]}}].



%\cite{Parker:2018yvk}
\bibitem{Parker:2018yvk}
D.~E.~Parker, X.~Cao, A.~Avdoshkin, T.~Scaffidi and E.~Altman,
``A Universal Operator Growth Hypothesis,''
Phys. Rev. X \textbf{9}, no.4, 041017 (2019)
doi:10.1103/PhysRevX.9.041017
[\href{https://arxiv.org/abs/1812.08657}{{\tt arXiv:1812.08657 [cond-mat.stat-mech]}}].

%138 citations counted in INSPIRE as of 12 Dec 2022
%\cite{Avdoshkin:2022xuw}
\bibitem{Avdoshkin:2022xuw}
A.~Avdoshkin, A.~Dymarsky and M.~Smolkin,
``Krylov complexity in quantum field theory, and beyond,''
[\href{https://arxiv.org/abs/2212.14429}{{\tt arXiv:2212.14429 [hep-th]}}].
%2 citations counted in INSPIRE as of 22 Feb 2023
%\cite{Camargo:2022rnt}
\bibitem{Camargo:2022rnt}
H.~A.~Camargo, V.~Jahnke, K.~Y.~Kim and M.~Nishida,
``Krylov Complexity in Free and Interacting Scalar Field Theories with Bounded Power Spectrum,''
[\href{https://arxiv.org/abs/2212.14702}{{\tt arXiv:2212.14702 [hep-th]}}].


%\cite{Dymarsky:2021bjq}
\bibitem{Dymarsky:2021bjq}
A.~Dymarsky and M.~Smolkin,
``Krylov complexity in conformal field theory,''
Phys. Rev. D \textbf{104}, no.8, L081702 (2021)
doi:10.1103/PhysRevD.104.L081702
[\href{https://arxiv.org/abs/2104.09514}{{\tt arXiv:2104.09514 [hep-th]}}].

%31 citations counted in INSPIRE as of 12 Dec 2022


%\cite{Dymarsky:2019elm}
\bibitem{Dymarsky:2019elm}
A.~Dymarsky and A.~Gorsky,
``Quantum chaos as delocalization in Krylov space,''
Phys. Rev. B \textbf{102}, no.8, 085137 (2020)
doi:10.1103/PhysRevB.102.085137
[\href{https://arxiv.org/abs/1912.12227}{{\tt arXiv:1912.12227 [cond-mat.stat-mech]}}].

%30 citations counted in INSPIRE as of 12 Dec 2022


%\cite{Avdoshkin:2019trj}
\bibitem{Avdoshkin:2019trj}
A.~Avdoshkin and A.~Dymarsky,
``Euclidean operator growth and quantum chaos,''
Phys. Rev. Res. \textbf{2}, no.4, 043234 (2020)
doi:10.1103/PhysRevResearch.2.043234
[\href{https://arxiv.org/abs/1911.09672}{{\tt arXiv:1911.09672 [cond-mat.stat-mech]}}].

%38 citations counted in INSPIRE as of 12 Dec 2022


%\cite{Barbon:2019wsy}
\bibitem{Barbon:2019wsy}
J.~L.~F.~Barb\'on, E.~Rabinovici, R.~Shir and R.~Sinha,
``On The Evolution Of Operator Complexity Beyond Scrambling,''
JHEP \textbf{10}, 264 (2019)
doi:10.1007/JHEP10(2019)264
[\href{https://arxiv.org/abs/1907.05393}{{\tt arXiv:1907.05393  [hep-th]}}].

%46 citations counted in INSPIRE as of 12 Dec 2022


%\cite{Khetrapal:2022dzy}
\bibitem{Khetrapal:2022dzy}
S.~Khetrapal,
``Chaos and operator growth in 2d CFT,''
[\href{https://arxiv.org/abs/2210.15860}{{\tt arXiv:2210.15860 [hep-th]}}].

%1 citations counted in INSPIRE as of 12 Dec 2022


%\cite{Rabinovici:2021qqt}
\bibitem{Rabinovici:2021qqt}
E.~Rabinovici, A.~S\'anchez-Garrido, R.~Shir and J.~Sonner,
``Krylov localization and suppression of complexity,''
JHEP \textbf{03}, 211 (2022)
doi:10.1007/JHEP03(2022)211
[\href{https://arxiv.org/abs/2112.12128}{{\tt arXiv:2112.12128  [hep-th]}}].

%15 citations counted in INSPIRE as of 12 Dec 2022


%\cite{Rabinovici:2022beu}
\bibitem{Rabinovici:2022beu}
E.~Rabinovici, A.~S\'anchez-Garrido, R.~Shir and J.~Sonner,
``Krylov complexity from integrability to chaos,''
JHEP \textbf{07}, 151 (2022)
doi:10.1007/JHEP07(2022)151
[\href{https://arxiv.org/abs/2207.07701}{{\tt arXiv:2207.07701  [hep-th]}}].

%11 citations counted in INSPIRE as of 12 Dec 2022


%5 citations counted in INSPIRE as of 12 Dec 2022
%\cite{Balasubramanian:2022tpr}
\bibitem{Balasubramanian:2022tpr}
V.~Balasubramanian, P.~Caputa, J.~M.~Magan and Q.~Wu,
``Quantum chaos and the complexity of spread of states,''
Phys. Rev. D \textbf{106}, no.4, 046007 (2022)
doi:10.1103/PhysRevD.106.046007
[\href{https://arxiv.org/abs/2202.06957}{{\tt arXiv:2202.06957 [hep-th]}}].

%\cite{Caputa:2022eye}
\bibitem{Caputa:2022eye}
P.~Caputa and S.~Liu,
``Quantum complexity and topological phases of matter,''
Phys. Rev. B \textbf{106}, no.19, 195125 (2022)
doi:10.1103/PhysRevB.106.195125
[\href{https://arxiv.org/abs/2205.05688}{{\tt arXiv:2205.05688 [hep-th]}}].


%\cite{Bhattacharjee:2022lzy}
\bibitem{Bhattacharjee:2022lzy}
B.~Bhattacharjee, X.~Cao, P.~Nandy and T.~Pathak,
``Operator growth in open quantum systems: lessons from the dissipative SYK,''
JHEP \textbf{03} (2023), 054
doi:10.1007/JHEP03(2023)054
[\href{https://arxiv.org/abs/2212.06180}{{\tt arXiv:2212.06180 [quant-ph]}}].


%26 citations counted in INSPIRE as of 12 Dec 2022
%\cite{Caputa:2022yju}
\bibitem{Caputa:2022yju}
P.~Caputa, N.~Gupta, S.~S.~Haque, S.~Liu, J.~Murugan and H.~J.~R.~Van Zyl,
``Spread Complexity and Topological Transitions in the Kitaev Chain,''
[\href{https://arxiv.org/abs/2208.06311}{{\tt arXiv:2208.06311 [hep-th]}}].

%11 citations counted in INSPIRE as of 12 Dec 2022
\bibitem{Afrasiar:2022efk}
M.~Afrasiar, J.~Kumar Basak, B.~Dey, K.~Pal and K.~Pal,
``Time evolution of spread complexity in quenched Lipkin-Meshkov-Glick model,''
[\href{https://arxiv.org/abs/2208.10520}{{\tt arXiv:2208.10520 [hep-th]}}].

%\cite{Alishahiha:2022nhe}
\bibitem{Alishahiha:2022nhe}
M.~Alishahiha
``On Quantum Complexity,''
[\href{https://arxiv.org/abs/2209.14689}{{\tt arXiv:2209.14689 [hep-th]}}].
%2 citations counted in INSPIRE as of 23 Mar 2023


%\cite{Jefferson:2017sdb}
\bibitem{Jefferson:2017sdb}
R.~Jefferson and R.~C.~Myers,
``Circuit complexity in quantum field theory,''
JHEP \textbf{10} (2017), 107
doi:10.1007/JHEP10(2017)107
[\href{https://arxiv.org/abs/1707.08570}{{\tt arXiv:1707.08570  [hep-th]}}].

%347 citations counted in INSPIRE as of 24 Mar 2023
%\cite{Chapman:2018hou}
\bibitem{Chapman:2018hou}
S.~Chapman, J.~Eisert, L.~Hackl, M.~P.~Heller, R.~Jefferson, H.~Marrochio and R.~C.~Myers,
``Complexity and entanglement for thermofield double states,''
SciPost Phys. \textbf{6}, no.3, 034 (2019)
doi:10.21468/SciPostPhys.6.3.034
[\href{https://arxiv.org/abs/1810.05151}{{\tt arXiv:1810.05151  [hep-th]}}].

%157 citations counted in INSPIRE as of 12 Dec 2022

%\cite{Caceres:2019pgf}
\bibitem{Caceres:2019pgf}
E.~Caceres, S.~Chapman, J.~D.~Couch, J.~P.~Hernandez, R.~C.~Myers and S.~M.~Ruan,
``Complexity of Mixed States in QFT and Holography,''
JHEP \textbf{03}, 012 (2020)
doi:10.1007/JHEP03(2020)012
[\href{https://arxiv.org/abs/1909.10557}{{\tt arXiv:1909.10557 [hep-th]}}].

%75 citations counted in INSPIRE as of 12 Dec 2022

%\cite{Chagnet:2021uvi}
\bibitem{Chagnet:2021uvi}
N.~Chagnet, S.~Chapman, J.~de Boer and C.~Zukowski,
``Complexity for Conformal Field Theories in General Dimensions,''
Phys. Rev. Lett. \textbf{128}, no.5, 051601 (2022)
doi:10.1103/PhysRevLett.128.051601
[\href{https://arxiv.org/abs/2103.06920}{{\tt arXiv:2103.06920 [hep-th]}}].

%42 citations counted in INSPIRE as of 12 Dec 2022




%\cite{Stanford:2014jda}
\bibitem{Stanford:2014jda}
D.~Stanford and L.~Susskind,
``Complexity and Shock Wave Geometries,''
Phys. Rev. D \textbf{90}, no.12, 126007 (2014)
doi:10.1103/PhysRevD.90.126007
[\href{https://arxiv.org/abs/1406.2678}{{\tt arXiv:1406.2678 [hep-th]}}].

%551 citations counted in INSPIRE as of 12 Dec 2022


%\cite{Susskind:2018pmk}
\bibitem{Susskind:2018pmk}
L.~Susskind,
``Three Lectures on Complexity and Black Holes,''
Springer, 2020,
ISBN 978-3-030-45108-0, 978-3-030-45109-7
doi:10.1007/978-3-030-45109-7
[\href{https://arxiv.org/abs/1810.11563}{{\tt arXiv:1810.11563 [hep-th]}}].

%113 citations counted in INSPIRE as of 12 Dec 2022


%\cite{Susskind:2018tei}
\bibitem{Susskind:2018tei}
L.~Susskind,
``Why do Things Fall?,''
[\href{https://arxiv.org/abs/1802.01198}{{\tt arXiv:1802.01198 [hep-th]}}].

%77 citations counted in INSPIRE as of 12 Dec 2022
%\cite{Chattopadhyay:2023fob}
\bibitem{Chattopadhyay:2023fob}
A.~Chattopadhyay, A.~Mitra and H.~J.~R.~van Zyl,
``Spread complexity as classical dilaton solutions,''
[\href{https://arxiv.org/abs/2302.10489}{{\tt arXiv:2302.10489[hep-th]}}].
%1 citations counted in INSPIRE as of 24 Mar 2023

%\cite{Chapman:2021jbh}
\bibitem{Chapman:2021jbh}
S.~Chapman and G.~Policastro,
``Quantum computational complexity from quantum information to black holes and back,''
Eur. Phys. J. C \textbf{82}, no.2, 128 (2022)
doi:10.1140/epjc/s10052-022-10037-1
[\href{https://arxiv.org/abs/2110.14672}{{\tt arXiv:2110.14672 [hep-th]}}].

%29 citations counted in INSPIRE as of 12 Dec 2022


%\cite{Caputa:2021ori}
\bibitem{Caputa:2021ori}
P.~Caputa and S.~Datta,
``Operator growth in 2d CFT,''
JHEP \textbf{12}, 188 (2021)
[erratum: JHEP \textbf{09}, 113 (2022)]
doi:10.1007/JHEP12(2021)188
[\href{https://arxiv.org/abs/2110.10519}{{\tt arXiv:2110.10519 [hep-th]}}].

%21 citations counted in INSPIRE as of 12 Dec 2022


%\cite{Belin:2021bga}
\bibitem{Belin:2021bga}
A.~Belin, R.~C.~Myers, S.~M.~Ruan, G.~S\'arosi and A.~J.~Speranza,
``Does Complexity Equal Anything?,''
Phys. Rev. Lett. \textbf{128}, no.8, 081602 (2022)
doi:10.1103/PhysRevLett.128.081602
[\href{https://arxiv.org/abs/2111.02429}{{\tt arXiv:2111.02429 [hep-th]}}].

%27 citations counted in INSPIRE as of 12 Dec 2022


%\cite{Kar:2021nbm}
\bibitem{Kar:2021nbm}
A.~Kar, L.~Lamprou, M.~Rozali and J.~Sully,
``Random matrix theory for complexity growth and black hole interiors,''
JHEP \textbf{01}, 016 (2022)
doi:10.1007/JHEP01(2022)016
[\href{https://arxiv.org/abs/2106.02046}{{\tt arXiv:2106.02046 [hep-th]}}].

%\cite{VMR:2008vsg}
\bibitem{VMR:2008vsg}
VS Viswanath and Gerhard M{\"u}llerller, 
``The Recursion Method: Application to Many Body Dynamics,''
(Springer, 2008)

%\cite{Fitzpatrick:2014vua}
\bibitem{Fitzpatrick:2014vua}
A.~L.~Fitzpatrick, J.~Kaplan and M.~T.~Walters,
``Universality of Long-Distance AdS Physics from the CFT Bootstrap,''
JHEP \textbf{08} (2014), 145
doi:10.1007/JHEP08(2014)145
[\href{https://arxiv.org/abs/1904.12819}{{\tt arXiv: 1403.6829 [hep-th]}}].

%\cite{Fitzpatrick:2015zha}
\bibitem{Fitzpatrick:2015zha}
A.~L.~Fitzpatrick, J.~Kaplan and M.~T.~Walters,
``Virasoro Conformal Blocks and Thermality from Classical Background Fields,''
JHEP \textbf{11} (2015), 200
doi:10.1007/JHEP11(2015)200
[\href{https://arxiv.org/abs/1904.12819}{{\tt arXiv:1501.05315 [hep-th]}}].

%\cite{Susskind:2019ddc}
\bibitem{Susskind:2019ddc}
L.~Susskind,
``Complexity and Newton's Laws,''
Front. in Phys. \textbf{8}, 262 (2020)
doi:10.3389/fphy.2020.00262
[\href{https://arxiv.org/abs/1904.12819}{{\tt arXiv:1904.12819 [hep-th]}}].

%47 citations counted in INSPIRE as of 02 Mar 2023

%\cite{Anous:2019yku}
\bibitem{Anous:2019yku}
T.~Anous and J.~Sonner,
``Phases of scrambling in eigenstates,''
SciPost Phys. \textbf{7}, 003 (2019)
doi:10.21468/SciPostPhys.7.1.003
[\href{https://arxiv.org/abs/1903.03143}{{\tt arXiv:1903.03143 [hep-th]}}].

%47 citations counted in INSPIRE as of 02 Mar 2023


%24 citations counted in INSPIRE as of 12 Dec 2022
%\cite{Kar:2021nbm}
\bibitem{Mattis:1987mat}
Michael P. Mattis,
``Correlations in 2-dimensional critical theories,''
Nuclear Physics B,Volume 285, 1987, Pages 671-686, ISSN 0550-3213,
[\href{https://doi.org/10.1016/0550-3213(87)90361-0}{{\tt doi: 10.1016/0550-3213(87)90361-0}}].

%\cite{Serbyn:2020wys}
\bibitem{Serbyn:2020wys}
M.~Serbyn, D.~A.~Abanin and Z.~Papi\'c,
``Quantum many-body scars and weak breaking of ergodicity,''
Nature Phys. \textbf{17}, no.6, 675-685 (2021)
doi:10.1038/s41567-021-01230-2
[\href{https://arxiv.org/abs/2011.0948}{{\tt arXiv:2011.0948 [quant-ph]}}].

%180 citations counted in INSPIRE as of 27 Feb 2023

%\cite{Moudgalya:2021xlu}
\bibitem{Moudgalya:2021xlu}
S.~Moudgalya, B.~A.~Bernevig and N.~Regnault,
``Quantum many-body scars and Hilbert space fragmentation: a review of exact results,''
Rept. Prog. Phys. \textbf{85}, no.8, 086501 (2022)
doi:10.1088/1361-6633/ac73a0
[\href{https://arxiv.org/abs/2109.00548}{{\tt arXiv:2109.00548 [cond-mat.str-el]}}].

%94 citations counted in INSPIRE as of 27 Feb 2023


%\cite{Banerjee:2020tgz}
\bibitem{Banerjee:2020tgz}
D.~Banerjee and A.~Sen,
``Quantum Scars from Zero Modes in an Abelian Lattice Gauge Theory on Ladders,''
Phys. Rev. Lett. \textbf{126}, no.22, 220601 (2021)
doi:10.1103/PhysRevLett.126.220601
[\href{https://arxiv.org/abs/2012.08540}{{\tt 2012.08540 [cond-mat.str-el]}}].
%44 citations counted in INSPIRE as of 19 Feb 2023


%\cite{Biswas:2022env}
\bibitem{Biswas:2022env}
S.~Biswas, D.~Banerjee and A.~Sen,
``Scars from protected zero modes and beyond in $U(1)$ quantum link and quantum dimer models,''
SciPost Phys. \textbf{12}, no.5, 148 (2022)
doi:10.21468/SciPostPhys.12.5.148
[\href{https://arxiv.org/abs/2202.03451}{{\tt 2202.03451 [cond-mat.str-el]}}].

%8 citations counted in INSPIRE as of 19 Feb 2023
%\cite{Bhattacharjee:2022qjw}
\bibitem{Bhattacharjee:2022qjw}
B.~Bhattacharjee, S.~Sur and P.~Nandy,
``Probing quantum scars and weak ergodicity breaking through quantum complexity,''
Phys. Rev. B \textbf{106} (2022) no.20, 205150
doi:10.1103/PhysRevB.106.205150
[\href{https://arxiv.org/abs/2208.05503}{{\tt 2208.05503 [quant-ph]}}].
%9 citations counted in INSPIRE as of 23 Mar 2023


%\cite{Alishahiha:2022anw}
\bibitem{Alishahiha:2022anw}
M.~Alishahiha and S.~Banerjee,
``A universal approach to Krylov State and Operator complexities,''
[\href{https://arxiv.org/abs/2212.10583}{{\tt arXiv:2212.10583 [hep-th]}}].
%4 citations counted in INSPIRE as of 23 Mar 2023


%\cite{Das:2021gts}
\bibitem{Das:2021gts}
D.~Das, R.~Ghosh and K.~Sengupta,
``Conformal Floquet dynamics with a continuous drive protocol,''
JHEP \textbf{05}, 172 (2021)
doi:10.1007/JHEP05(2021)172
[\href{https://arxiv.org/abs/2101.04140}{{\tt arXiv:2101.04140 [hep-th]}}].

%9 citations counted in INSPIRE as of 19 Feb 2023


%\cite{Wen:2021mlv}
\bibitem{Wen:2021mlv}
X.~Wen, Y.~Gu, A.~Vishwanath and R.~Fan,
``Periodically, Quasi-periodically, and Randomly Driven Conformal Field Theories (II): Furstenberg's Theorem and Exceptions to Heating Phases,''
SciPost Phys. \textbf{13}, no.4, 082 (2022)
doi:10.21468/SciPostPhys.13.4.082
[\href{https://arxiv.org/abs/2109.10923}{{\tt arXiv:2109.10923  [cond-mat.stat-mech]}}].

%2 citations counted in INSPIRE as of 19 Feb 2023


%\cite{Fan:2020orx}
\bibitem{Fan:2020orx}
R.~Fan, Y.~Gu, A.~Vishwanath and X.~Wen,
``Floquet conformal field theories with generally deformed Hamiltonians,''
SciPost Phys. \textbf{10}, no.2, 049 (2021)
doi:10.21468/SciPostPhys.10.2.049
[\href{https://arxiv.org/abs/2011.09491}{{\tt arXiv:2011.09491  [hep-th]}}].

%14 citations counted in INSPIRE as of 19 Feb 2023


%\cite{Wen:2020wee}
\bibitem{Wen:2020wee}
X.~Wen, R.~Fan, A.~Vishwanath and Y.~Gu,
``Periodically, quasiperiodically, and randomly driven conformal field theories,''
Phys. Rev. Res. \textbf{3}, no.2, 023044 (2021)
doi:10.1103/PhysRevResearch.3.023044
[\href{https://arxiv.org/abs/2006.10072}{{\tt arXiv:2006.10072 [cond-mat.stat-mech]}}].

%22 citations counted in INSPIRE as of 19 Feb 2023


%\cite{Das:2022pez}
\bibitem{Das:2022pez}
S.~Das, B.~Ezhuthachan, A.~Kundu, S.~Porey, B.~Roy and K.~Sengupta,
``Brane Detectors of a Dynamical Phase Transition in a Driven CFT,''
[\href{https://arxiv.org/abs/2212.04201}{{\tt arXiv:2212.04201  [hep-th]}}].

%1 citations counted in INSPIRE as of 19 Feb 2023


%\cite{Das:2022jrr}
\bibitem{Das:2022jrr}
S.~Das, B.~Ezhuthachan, A.~Kundu, S.~Porey, B.~Roy and K.~Sengupta,
``Out-of-Time-Order correlators in driven conformal field theories,''
JHEP \textbf{08}, 221 (2022)
doi:10.1007/JHEP08(2022)221
[\href{https://arxiv.org/abs/2202.12815}{{\tt arXiv:2202.12815  [hep-th]}}].
%6 citations counted in INSPIRE as of 19 Feb 2023

%\cite{Caputa:2022zsr}
\bibitem{Caputa:2022zsr}
P.~Caputa and D.~Ge,
``Entanglement and geometry from subalgebras of the Virasoro,''
[\href{https://arxiv.org/abs/2211.03630}{{\tt arXiv:2211.03630  [hep-th]}}].

%5 citations counted in INSPIRE as of 27 Feb 2023


%\cite{Maldacena:2015waa}
\bibitem{Maldacena:2015waa}
J.~Maldacena, S.~H.~Shenker and D.~Stanford,
``A bound on chaos,''
JHEP \textbf{08}, 106 (2016)
doi:10.1007/JHEP08(2016)106
[\href{https://arxiv.org/abs/1503.01409}{{\tt arXiv:1503.01409 [hep-th]}}].

%1471 citations counted in INSPIRE as of 19 Feb 2023


%\cite{Banerjee:2019vff}
\bibitem{Banerjee:2019vff}
A.~Banerjee, A.~Kundu and R.~R.~Poojary,
``Rotating black holes in AdS spacetime, extremality, and chaos,''
Phys. Rev. D \textbf{102}, no.10, 106013 (2020)
doi:10.1103/PhysRevD.102.106013
[\href{https://arxiv.org/abs/1912.12996}{{\tt arXiv:1912.12996  [hep-th]}}].

%17 citations counted in INSPIRE as of 19 Feb 2023


%\cite{Banerjee:2018kwy}
\bibitem{Banerjee:2018kwy}
A.~Banerjee, A.~Kundu and R.~Poojary,
``Maximal Chaos from Strings, Branes and Schwarzian Action,''
JHEP \textbf{06}, 076 (2019)
doi:10.1007/JHEP06(2019)076
[\href{https://arxiv.org/abs/1811.04977}{{\tt arXiv:1811.04977   [hep-th]}}].

%20 citations counted in INSPIRE as of 19 Feb 2023


%\cite{Banerjee:2018twd}
\bibitem{Banerjee:2018twd}
A.~Banerjee, A.~Kundu and R.~R.~Poojary,
``Strings, branes, Schwarzian action and maximal chaos,''
Phys. Lett. B \textbf{838}, 137632 (2023)
doi:10.1016/j.physletb.2022.137632
[\href{https://arxiv.org/abs/1809.02090}{{\tt arXiv:1809.02090  [hep-th]}}].

%27 citations counted in INSPIRE as of 19 Feb 2023

%\cite{Malvimat:2021itk}
\bibitem{Malvimat:2021itk}
V.~Malvimat and R.~R.~Poojary,
``Fast scrambling due to rotating shockwaves in BTZ,''
Phys. Rev. D \textbf{105} (2022) no.12, 126019
doi:10.1103/PhysRevD.105.126019
[\href{https://arxiv.org/abs/2112.14089}{{\tt arXiv:2112.14089 [hep-th]}}].

%\cite{Malvimat:2022oue}
\bibitem{Malvimat:2022oue}
V.~Malvimat and R.~R.~Poojary,
``Fast scrambling of mutual information in Kerr-AdS$_{\textbf{4}}$ spacetime,''
Phys. Rev. D \textbf{107} (2023) no.2, 026019
doi:10.1103/PhysRevD.107.026019
[\href{https://arxiv.org/abs/2207.13022}{{\tt arXiv:2207.13022 [hep-th]}}].
%4 citations counted in INSPIRE as of 21 Feb 2023

%\cite{Malvimat:2022fhd}
\bibitem{Malvimat:2022fhd}
V.~Malvimat and R.~R.~Poojary,
``Fast Scrambling of mutual information in Kerr-AdS$_{\textbf{5}}$,''
[\href{https://arxiv.org/abs/2210.02950}{{\tt arXiv:2210.02950 [hep-th]}}].


%\cite{Gao:2016bin}
\bibitem{Gao:2016bin}
P.~Gao, D.~L.~Jafferis and A.~C.~Wall,
``Traversable Wormholes via a Double Trace Deformation,''
JHEP \textbf{12}, 151 (2017)
doi:10.1007/JHEP12(2017)151
[\href{https://arxiv.org/abs/1608.05687}{{\tt arXiv:1608.05687  [hep-th]}}].

%325 citations counted in INSPIRE as of 19 Feb 2023


%\cite{Maldacena:2017axo}
\bibitem{Maldacena:2017axo}
J.~Maldacena, D.~Stanford and Z.~Yang,
``Diving into traversable wormholes,''
Fortsch. Phys. \textbf{65}, no.5, 1700034 (2017)
doi:10.1002/prop.201700034
[\href{https://arxiv.org/abs/1704.05333}{{\tt arXiv:1704.05333  [hep-th]}}].

%300 citations counted in INSPIRE as of 19 Feb 2023


%\cite{Kundu:2021nwp}
\bibitem{Kundu:2021nwp}
A.~Kundu,
``Wormholes and holography: an introduction,''
Eur. Phys. J. C \textbf{82}, no.5, 447 (2022)
doi:10.1140/epjc/s10052-022-10376-z
[\href{https://arxiv.org/abs/2110.14958}{{\tt arXiv:2110.14958  [hep-th]}}].

%\cite{Lashkari:2016vgj}
\bibitem{Lashkari:2016vgj}
N.~Lashkari, A.~Dymarsky and H.~Liu,
``Eigenstate Thermalization Hypothesis in Conformal Field Theory,''
J. Stat. Mech. \textbf{1803}, no.3, 033101 (2018)
doi:10.1088/1742-5468/aab020
[\href{https://arxiv.org/abs/1610.00302}{{\tt arXiv:1610.00302 [hep-th]}}].

%89 citations counted in INSPIRE as of 04 Mar 2023

%\cite{Banerjee:2016qca}
\bibitem{Banerjee:2016qca}
P.~Banerjee, S.~Datta and R.~Sinha,
``Higher-point conformal blocks and entanglement entropy in heavy states,''
JHEP \textbf{05}, 127 (2016)
doi:10.1007/JHEP05(2016)127
[\href{https://arxiv.org/abs/1601.06794}{{\tt arXiv:1601.06794  [hep-th]}}].

%57 citations counted in INSPIRE as of 04 Mar 2023

%\cite{Rabinovici:2023yex}
\bibitem{Rabinovici:2023yex}
E.~Rabinovici, A.~S\'anchez-Garrido, R.~Shir and J.~Sonner,
``A bulk manifestation of Krylov complexity,''
[\href{https://arxiv.org/abs/arXiv:2305.04355}{{\tt arXiv:arXiv:2305.04355 [hep-th]}}].
%9 citations counted in INSPIRE as of 22 Jul 2023

%\cite{Muck:2022xfc}
\bibitem{Muck:2022xfc}
W.~M\"uck and Y.~Yang,
``Krylov complexity and orthogonal polynomials,''
Nucl. Phys. B \textbf{984}, 115948 (2022)
doi:10.1016/j.nuclphysb.2022.115948
[\href{https://arxiv.org/abs/arXiv:2305.04355}{{\tt arXiv:2205.12815 [hep-th]}}].

%20 citations counted in INSPIRE as of 22 Jul 2023

\end{thebibliography}
\end{document}